\documentclass[
 reprint,
superscriptaddress,
bibnotes,
 amsmath,amssymb,
 aps,
 prl,
]{revtex4-2}

\usepackage{graphicx}
\usepackage{dcolumn}
\usepackage{bm}
\usepackage{pdfcomment}
\usepackage{svg}
\usepackage{float}
\usepackage{amsthm}

\usepackage{xcolor}
\usepackage{soul}

\newcommand{\bX}{\textbf{X}}

\newcommand{\tr}{\mathrm{tr}}

\usepackage{bibunits}
\defaultbibliographystyle{apsrev4-2}
\defaultbibliography{references}

\begin{document}

\begin{bibunit}

\title{Synergistic Motifs in Gaussian Systems}

\author{Enrico Caprioglio}
 \email{Contact author: e.caprioglio@sussex.ac.uk}
 \affiliation{%
  Department of Informatics, University of Sussex, Brighton, United Kingdom
  }%
\author{Pedro A. M. Mediano}
  \affiliation{%
  Department of Computing, Imperial College London, London, United Kingdom}
  \affiliation{%
  Division of Psychology and Language Sciences, University College London, London, United Kingdom
  }
\author{Luc Berthouze}%
  \affiliation{%
  Department of Informatics, University of Sussex, Brighton, United Kingdom
  }%

\date{\today}

\begin{abstract}
High-order interdependencies are central features of complex systems, yet a mechanistic explanation for their emergence remains elusive. Currently, it is unknown under what conditions high-order interdependencies, quantified by the information-theoretic construct of synergy, arise in systems governed by pairwise interactions. We solve this problem by providing precise sufficient and necessary conditions for when synergy prevails over low-order interdependencies in the weak interaction regime, namely, we prove that antibalanced (highly frustrated) correlational structures in Gaussian systems are sufficient for synergy-dominance and that antibalanced interaction motifs in Ornstein–Uhlenbeck processes are necessary for synergy-dominance. We validate the applicability of these analytical insights beyond the weak interaction regime, as well as in Ising, oscillatory, and empirical networks from multiple domains. Our results demonstrate that pairwise interactions can give rise to synergistic information in the absence of explicit high-order mechanisms, and highlight structural balance theory as an instrumental conceptual framework to study high-order interdependencies.

\end{abstract}

\maketitle

\textit{Introduction.}---Complex systems are characterized by collective behaviours that cannot be reduced to their components' individual properties. The investigation of the mechanisms that can give rise to these behaviours is at the heart of complexity science~\cite{Rosenblueth1945, Jensen2022, Rosas2022}, a central problem across domains of physics~\cite{Battiston2021,Milln2025,nijholt2022emergent}, and a fundamental question in neuroscience~\cite{Sporns2022}, genetics~\cite{Park2021}, ecology~\cite{Levine2017} and evolutionary biology~\cite{Carroll2001, Krakauer2020}.

Recent work has increasingly focused on the intuitive relationship between high-order mechanisms (i.e., beyond pairwise interaction terms in the system's generative process, such as a Hamiltonian) and collective behaviours (i.e., beyond pairwise statistical dependencies in the system's resulting activity)~\cite{Rosas2022, Battiston2020, Battiston2021, Milln2025}. Advances in information theory---formalizing high-order interdependencies through the notion of synergy~\cite{Williams2010, IncePED}---have enabled the systematic study of the interplay between mechanisms and high-order behaviours \cite{Rosas2019, Robiglio2025}. Surprisingly, while high-order interdependencies are enhanced by high-order mechanisms~\cite{Rosas2019, Robiglio2025}, they can also arise purely from low-order mechanisms~\cite{Matsuda2000, Barrett2015}. This puzzle motivates the central question we address here: under what precise minimal conditions can low-order mechanisms produce synergistic behaviours?

Early studies in two restricted settings provide some insights. For Ising models of size $N=3$, frustrated pairwise spin couplings lead to a negative interaction information~\cite{Matsuda2000}, now understood to be indicative of synergistic interdependencies~\cite{Williams2010}. In Gaussian systems of size $N=3$, systems are synergistic if two variables both correlate with a third but are anti- or un-correlated with each other~\cite{Barrett2015, Kontoyiannis2005, Liardi2024}. However, a formal proof linking correlation and anticorrelation patterns to synergy-dominance is missing, and whether this intuition translates to larger systems remains unknown.

To address these open problems, we combine two tools: Structural balance theory (SBT) and the O-information ($\Omega$). SBT~\cite{Heider1946} provides a graph-theoretic framework~\cite{Harary1953, Cartwright1956} to study patterns of signed pairwise relationships across domains~\cite{Healy1973, Maoz2010, Facchetti2011, Bramson2017,  Estrada2019,  Masoomy2021, Moradimanesh2021, Saberi2021, Talesh2023, Saberi2024}. In spin glass theory, balanced graphs minimize spins' frustration, while antibalanced graphs maximise it~\cite{Antal2005, Marvel2009, Rabbani2019, Malarz2022, Pham2022}. The O-information is a high-order extension of the mutual information~\cite{Rosas2019} whose sign captures the relative strength of synergistic (high-order) vs redundant (low-order) interdependencies in a system~\cite{Rosas2024}. Importantly, it is analytically tractable and readily computable, thus serving as the ideal aggregate measure of high-order interdependencies in large multivariate systems \cite{Gatica2021, Varley2023C, Varley2023, Puxeddu2025, Varley2025}.

In this letter, we provide precise conditions under which synergistic interdependencies arise from pairwise mechanisms. For static Gaussian systems, we derive a condition in terms of signed patterns in the correlation matrix and prove that, in any dimension, antibalanced structures ensure synergy-dominance. Extending this to dynamical systems, we prove that antibalanced interaction motifs in Ornstein–Uhlenbeck processes are necessary for synergy-dominance. Finally, we demonstrate the relevance of our results across domains of physics by numerically showing that antibalanced motifs are predictors of synergy-dominance in systems taken from statistical mechanics and electrochemical oscillators, as well as in medical, financial, physiological, and epidemiological empirical datasets.

\textit{Notation and preliminaries.}---Let $G$ be a signed graph with adjacency matrix $W\in\mathbb{R}^{N\times N}$, $W_{ii}=0$. Harary's structure theorem~\cite{Harary2007} states that a graph $G$ is \emph{balanced} (\emph{antibalanced}) if it can be partitioned into two non-overlapping sets of vertices such that edges between vertices in the same set are positive (negative) and edges between nodes in distinct sets are negative (positive). Let $\boldsymbol{w}^k$ be the set of all closed walks $w$ of length $k$ in $G$, and for $w = (i_1,i_2,\dots,i_k,i_1)\in\boldsymbol{w}^k$ define the weight of a walk as $\sigma(w)=\prod_{i=1}^kW_{w_i,w_{i+1}}$. This theorem implies that if $G$ is \emph{antibalanced} then every closed walk of length $k$ satisfies $\mathrm{sign}(\sigma(w))=(-1)^k$.

Let $\bX=\{X_1,X_2,\dots,X_N\}$ be a multivariate process of size $N$ and let $\bX_{-i} = \bX\setminus{X_i}$ denote the same system with the $i$\textsuperscript{th} variable removed. The O-information can be written as~\cite{Varley2023}:
\begin{align}
    \Omega(\bX) = \sum_{i=1}^N TC(\bX_{-i}) - (N-2) TC(\bX), \label{eqn:o-info-only-tc}
\end{align}
where the total correlation $TC(\bX)=\sum_jH(X_j)-H(\bX)$~\cite{Watanabe1960} quantifies the strength of correlations within the system as a whole~\cite{Rosas2019, Varley2023}. A negative sign ($\Omega<0$) indicates that high-order interdependencies dominate (synergy-dominance); a positive sign ($\Omega>0$) instead indicates that low-order interdependencies dominate (redundancy-dominance)~\cite{Rosas2019, Varley2023C, Rosas2024}.

\textit{Static Gaussian systems.}---Information-theoretic quantities can be studied analytically in Gaussian systems, allowing us to derive precise conditions for synergy-dominance. In this regime, the O-information can be conveniently written in terms of blocks of the covariance matrix as
\begin{align}
    \Omega(\bX) = \frac{N-2}{2}\log\Big[\det(\Sigma)\Big] - \frac{1}{2}\sum_i\log\Big[\det(\Sigma_{-i})\Big],
    \label{eqn:o-gaussian}
\end{align}
where $\Sigma_{-i}$ indicates the system's covariance matrix with the $i$\textsuperscript{th} variable removed (see Supplemental Material (SM)~\cite{SM}~\nocite{Cover2005, Neri2024, Koseska2013, Hens2013, Luppi2024B, Lizier2014} for step-by-step derivations). Without loss of generality, in what follows we consider standardized covariances $\Sigma$ (i.e., unit-variance correlation matrices). To study the signed patterns in $\Sigma$ using SBT, we represent the correlational structure of Gaussian systems as a \emph{complete signed} graph $G$ with adjacency matrix $W=\Sigma - \mathbb{I}$, where $\mathbb{I}$ is the identity matrix. We simplify Eq.~\eqref{eqn:o-gaussian}, following a similar approach as in \cite{Barnett2009}, by expanding the log determinants using the Mercator series
\begin{equation}
    \log\big[\det(\Sigma)\big] = - \sum_{k=2}\frac{(-1)^k}{k}\tr[W^k].
    \label{eqn:log-det-approximation}
\end{equation}
Importantly, this approximation converges only if the spectral radius of $W$, denoted as $\rho(W)$, is less than one. As noted in~\cite{Barnett2009}, this applies for the case of weakly coupled systems, which are common across the complex systems literature~\cite{Berlow1999, Csermely2004, Garas2008}. Using the trace identity $\sum_i^N\mathrm{tr}[W_{-i}^k] = (N-k)\mathrm{tr}[W^k]$ for $k=2,3$ (see~\cite{SM}) we can use Eq. \eqref{eqn:log-det-approximation} to isolate the lowest order term in $\Omega$:
\begin{equation}
    \Omega(\bX) = \frac16 \tr[W^3] + R_4(W). \label{eqn:first-approximation}
\end{equation}
where $R_4(W)\leq\mathcal{O}(\rho(W)^4)$ as $\rho(W)\to0$.

This approximation yields an important insight: Since $\tr[W^3]=\sum_{w\in\boldsymbol{w}^3}\sigma(w)$, whenever $|\frac16\tr[W^3]|>|R_4(W)|$ we have that $\bX$ is synergy-dominated if the triangles (i.e., length-3 walks) are antibalanced. In fact, the lowest order term in Eq.~\eqref{eqn:first-approximation} is, up to normalization and sign, equivalent to the structural balance-energy $U$~\cite{Marvel2009}
\begin{equation}
    U= -\frac{1}{\binom{N}{3}}\sum_{\{i,j,k\}}{W_{ij}W_{jk}W_{ik}},\label{eqn:structural-energy}
\end{equation}
a measure of balance that is minimized (maximised) if each triangle $\{i,j,k\}$ in the network is balanced (antibalanced). In what follows we study the higher-order corrections in Eq.~(\ref{eqn:first-approximation}) and use SBT to provide a stronger condition, proving that systems with antibalanced correlational structures must be synergistic. To do this, we introduce a trace identity valid for any $k$.

\begin{figure}[t]
    \centering
    \includegraphics[width=1\linewidth]{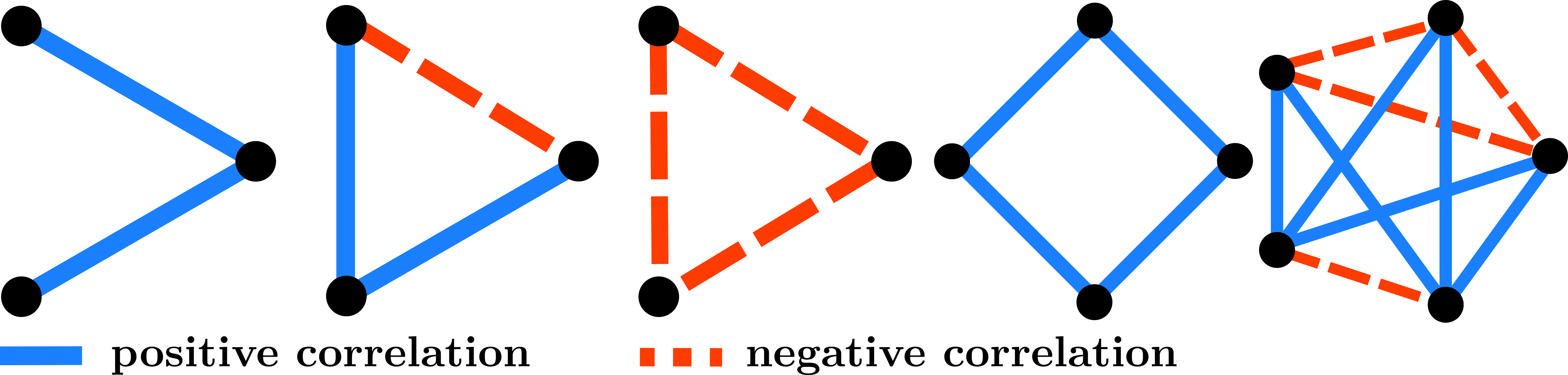}
    \caption{\textbf{Examples of \emph{synergistic motifs} of pairwise correlations that ensure synergy-dominance}. These structures are either antibalanced or only contain closed walks $w$ of even length with positive weight $\sigma(w)$. Missing edges indicate uncorrelated relationships.}
    \label{fig:example-syn-dominant-corr-structures}
\end{figure}

\textit{Proposition 1}---For $k\in\mathbb{N}$, $k>0$, we have
\begin{align*}
    \sum_i\tr[W_{-i}^k] = N\tr[W^k] - \sum_{w\in\boldsymbol{w}^k}|w|\sigma(w),
\end{align*}
where $|w|$ is the number of unique nodes encountered along the closed walk $w$. See proof in the SM~\cite{SM}.

We use this identity to eliminate the explicit dependence on $W_{-i}$ in the higher-order terms of Eq.~\eqref{eqn:first-approximation}, and derive the key theorem for static systems:

\textit{Theorem 1}---If $G$ is antibalanced, then $\Omega(\bX)<0$.

To prove this, insert Eq.~\eqref{eqn:log-det-approximation} into Eq.~\eqref{eqn:o-gaussian} and apply Proposition~1 to obtain
\begin{align}
    \Omega(\bX) &= \sum_{k=3}^\infty\Bigg[\frac{(-1)^{k-1}}{2k}\sum_{w\in\boldsymbol{w}^k}(|w|-2)\sigma(w)\Bigg]. \label{eqn:O-information-final-walk-expansion}
\end{align}
If $G$ is antibalanced, every nonzero closed walk $w\in\boldsymbol{w}^k$ in $G$ has $(-1)^{k-1}\sigma(w)<0$, and therefore $\Omega < 0$.

This result furnishes an exact sufficient condition for synergy-dominance and allows us to identify the signed patterns in the correlation matrix (i.e., closed walks $w$ in $W$) which contribute synergistically. The sign of each summation term in Eq.~\eqref{eqn:O-information-final-walk-expansion} depends on both the parity of $k$ and the sign of the walk's weight $\sigma(w)$. The \emph{synergistic motifs}, i.e., the closed walks that contribute negatively in Eq.~\eqref{eqn:O-information-final-walk-expansion}, thus correspond to even-length closed walks with $\sigma(w)>0$ and odd-length closed walks with $\sigma(w)<0$ (see examples in Fig. \ref{fig:example-syn-dominant-corr-structures}).

Notably, while our proof assumes $\rho(W)<1$, our extensive numerical analyses (see \textit{Experimental validation} section and SM~\cite{SM}) have not found any redundancy-dominated antibalanced correlational structures even when $\rho(W)\geq1$.

\textit{Dynamical Gaussian systems.}---We now ask which patterns of interactions in the coupling matrix of a linear dynamical system can lead to synergistic correlational structures. For example, in diffusive processes with only positive couplings, two variables that both correlate positively with a third tend to be correlated~\cite{Granovetter1973WeakTies}. Synergy-dominated systems instead (Fig.~\ref{fig:example-syn-dominant-corr-structures}) break this and hence the un- or anti-correlated pair must be negatively coupled. We therefore conjecture that antibalanced interaction motifs are required for dynamical Gaussian processes to be synergy-dominated.

Following a similar approach as Ref.~\cite{Barnett2009}, we consider Ornstein–Uhlenbeck (OU) processes: $d\bX(t) = -\bX(t)\cdot(\mathbb{I} - A)dt + d\boldsymbol{\mathcal{W}}(t)$, where $A\in\mathbb{R}^{N\times N}$ is the interaction (or coupling) matrix and $\boldsymbol{\mathcal{W}}$ is a Wiener process with identity covariance. For symmetric, Schur stable $A$, the stationary covariance can be written as (see~\cite{Barnett2009, Novelli2020, Lizier2023})
\begin{equation}
    \tilde{\Sigma} = \frac12(\mathbb{I} - A)^{-1}.\label{eqn:OU-solution}
\end{equation}
Then, by normalizing $\tilde{\Sigma}$ we obtain the correlation matrix $\Sigma$. With this closed-form expression, we prove our conjecture exactly for $N=3$ and in the weak interaction regime for arbitrary $N$. More precisely, we call weak interaction the regime where $\rho(A)\ll1$ (implying $\rho(W)<1$), and strong interaction the regime with $0<\rho(A)<1$ where the system is still Schur-stable (so information-theoretic quantities are well defined) but correlations are strong and $\rho(W)>1$ is allowed.

\textit{Theorem 2}---Let $\bX$ be a stable OU process with $A\in\mathbb{R}^{3\times3}$. If $\Omega(\bX)<0$, then $A$ is antibalanced.

To derive this, let $a=\Sigma_{12},b=\Sigma_{13},c=\Sigma_{23}$. From Eq.~\eqref{eqn:o-gaussian}, the condition $\Omega<0$ simplifies to $2abc < (ab)^2+(ac)^2+(bc)^2 - (abc)^2$. To connect this to the interactions $A_{ij}$, we use Eq.~\eqref{eqn:OU-solution} to write the same condition as $-2xyz > (xy)^2 + (xz)^2 + (yz)^2 - (xyz)^2$, where $x=A_{12}, y=A_{13}, z=A_{23}$ (see SM~\cite{SM}). For Schur-stable $A$, the RHS is always positive, so this expression is satisfied only if $xyz<0$. Thus, antibalanced interactions are \emph{necessary} for synergy-dominance when $N=3$.

For systems of arbitrary dimension, elements in $\tilde\Sigma$ are more complex. In the weak interaction regime, Eq.~\eqref{eqn:first-approximation} is dominated by the cubic term when $A$ is balanced. The SM~\cite{SM} provides analytical bounds on $R_4(W)$ in the balanced case, and numerical tests to identify the range of $\rho(W)$ for which $|\frac16 \tr[W^3]| > |R_4(W)|$. Under this condition, we prove that antibalanced interaction motifs are necessary for the system to be synergy-dominated by studying the sign relationships between $A$ and $W$.

\textit{Theorem 3}---Let $\bX$ be a stable OU process with $A\in\mathbb{R}^{N\times{N}}$. In the weak interaction regime, antibalanced interaction motifs in $A$ are necessary for $\Omega(\bX)<0$.

To prove this, we first show in the weak interaction limit that if $A$ is balanced then $\Omega(\bX)>0$. For Schur stable $A$, the Neumann series of Eq.~\eqref{eqn:OU-solution}, after normalization, yields
\begin{align}
    W_{ij}=\frac12\frac{\sum_{l=1}^{\infty}(A^l)_{ij}}{\sqrt{d_id_j}}, \quad i\neq{j},\;d_i=(\tilde\Sigma)_{ii}>0\,.\label{eqn:Proposition 2}
\end{align}
The denominator is positive, so the sign of $W_{ij}$ is the same as that of $\sum_l(A^l)_{ij}$ i.e., the contribution of all length-$l$ interaction walks from $i$ to $j$. If $A$ is balanced, Harary’s structure theorem for balance guarantees that any walk from node $i$ to $j$ satisfies $(A^l)_{ij}>0$ if $i$ and $j$ are in the same set, and $(A^l)_{ij}<0$ otherwise. Consequently, $\mathrm{sign}\left(\sum_{l}(A^{l})_{ij}\right)=\mathrm{sign}(A_{ij})$, which implies that $W_{ij}$ and $A_{ij}$ must share the same sign. In the weak interaction regime $|\frac16\tr[W^3]|>|R_4(W)|$, such that $\mathrm{sign}(\Omega(\bX))=\mathrm{sign}(\tr[W^3])$. Thus, for balanced $A$ we have $\mathrm{sign}(\Omega(\bX))=\mathrm{sign}(\tr[A^3])>0$ and the stationary dynamics of the resulting OU process is ensured to be redundancy-dominated. Consequently, we conclude that antibalanced interaction motifs (e.g., antibalanced triangles of interactions), which leave $A$ \emph{not} balanced, are \emph{necessary} for the system to be synergy-dominated.

\begin{figure}[t]
    \centering
    \includegraphics[width=1\linewidth]{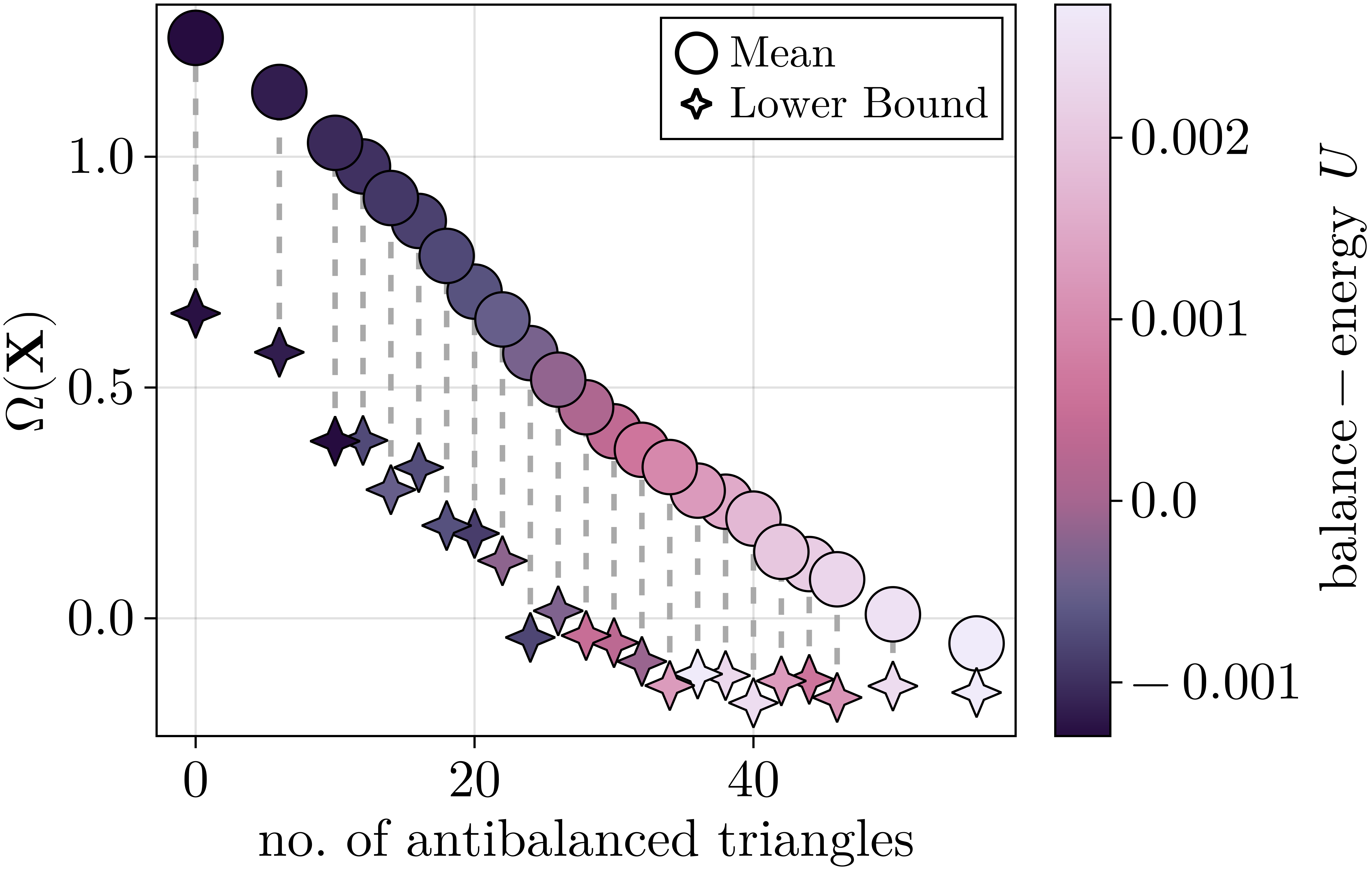}
     \caption{\textbf{Antibalanced interaction motifs are necessary for synergy-dominance in dynamical Gaussian systems with strong interactions}. Mean $\Omega(\boldsymbol{X})$ as a function of the number of antibalanced triangles in the interaction matrix. Lower bounds indicate the lowest $\Omega(\bX)$ encountered in our numerical exploration across all configurations with a specific no. of antibalanced triangles. Each mean $\Omega(\bX)$ is coloured according to its mean balance-energy value (Eq. \eqref{eqn:structural-energy}). Lower bounds are coloured with the energy value of the configuration that resulted in the lowest recorded $\Omega(\bX)$, not the highest recorded energy (not shown). Each $A\in\mathbb{R}^{8\times8}$ has $\rho(A) \approx 0.88$ (see SM~\cite{SM}).}
    \label{fig:N-8-OU-process-results}
\end{figure}

To test our conjecture beyond the weak interaction limit, we performed extensive numerical explorations for systems of size $N\leq9$ with varied interaction strengths $\rho(A)$ which ensure $\rho(W)>1$. Since the O-information is permutation-invariant \cite{Rosas2019}, we explored all \emph{non-isomorphic complete signed graphs}, each one encoding a unique sign pattern in $A$. The restriction to complete graphs allowed us to characterize each network of interactions by the number of antibalanced triangles only---balance is fully determined by triangles in this case~\cite{Antal2005}---and thus compare each graph on equal terms to identify the most synergistic structures. For non-complete graphs a similar systematic analysis is not yet obvious due to, for example, degree correlations, which is an interesting direction for future work.

We present our results in Fig. \ref{fig:N-8-OU-process-results} for $N=8$. For all system sizes studied (see SM~\cite{SM}), we found that the O-information is strongly anticorrelated with the number of antibalanced triangles in $A$. Importantly, balanced interaction structures (with $0$ antibalanced triangles) always result in redundancy-dominated systems, while antibalanced structures (with $\binom{N}{3}$ antibalanced triangles) consistently result in systems with the lowest mean $\Omega$.

\textit{Experimental validation.}---Gaussian systems often serve as analytically tractable starting point to analyse complex models~\cite{Barnett2009, Barrett2015}. We now show that the same relationship between antibalanced interactions and synergistic interdependencies extends to two prototypical models of complex systems: the landmark Ising model (see e.g.~\cite{Jensen2022}) and networks of subcritical Stuart-Landau (SL) oscillators, used widely in the modelling of electrochemical and neuronal systems~\cite{bisquert2022hopf}. Additionally, in empirical studies of continuous, high-dimensional data (such as fMRI, biochemical systems, or financial time series), information-theoretic quantities are most commonly estimated using Gaussian (or Gaussian-copula) models~\cite{Chen2003, Hlinka2011, Ince2016, Nozari2023}. Motivated by the widespread use of this approach, (see e.g.,~\cite{Tononi1994, Barnett2009, Tostevin2010, Hahn2023, Kringelbach2023, Scagliarini2022, Varley2025, Varley2023C, Puxeddu2025, Varley2023, Garas2008, Castro2025, Luppi2022, Luppi2024, Rajpal2025}) we demonstrate the practical relevance of our analytical insights (Eq.~\eqref{eqn:first-approximation} and Theorem 1) on financial and fMRI brain activity data.

\begin{figure}
    \centering
    \includegraphics[width=1\linewidth]{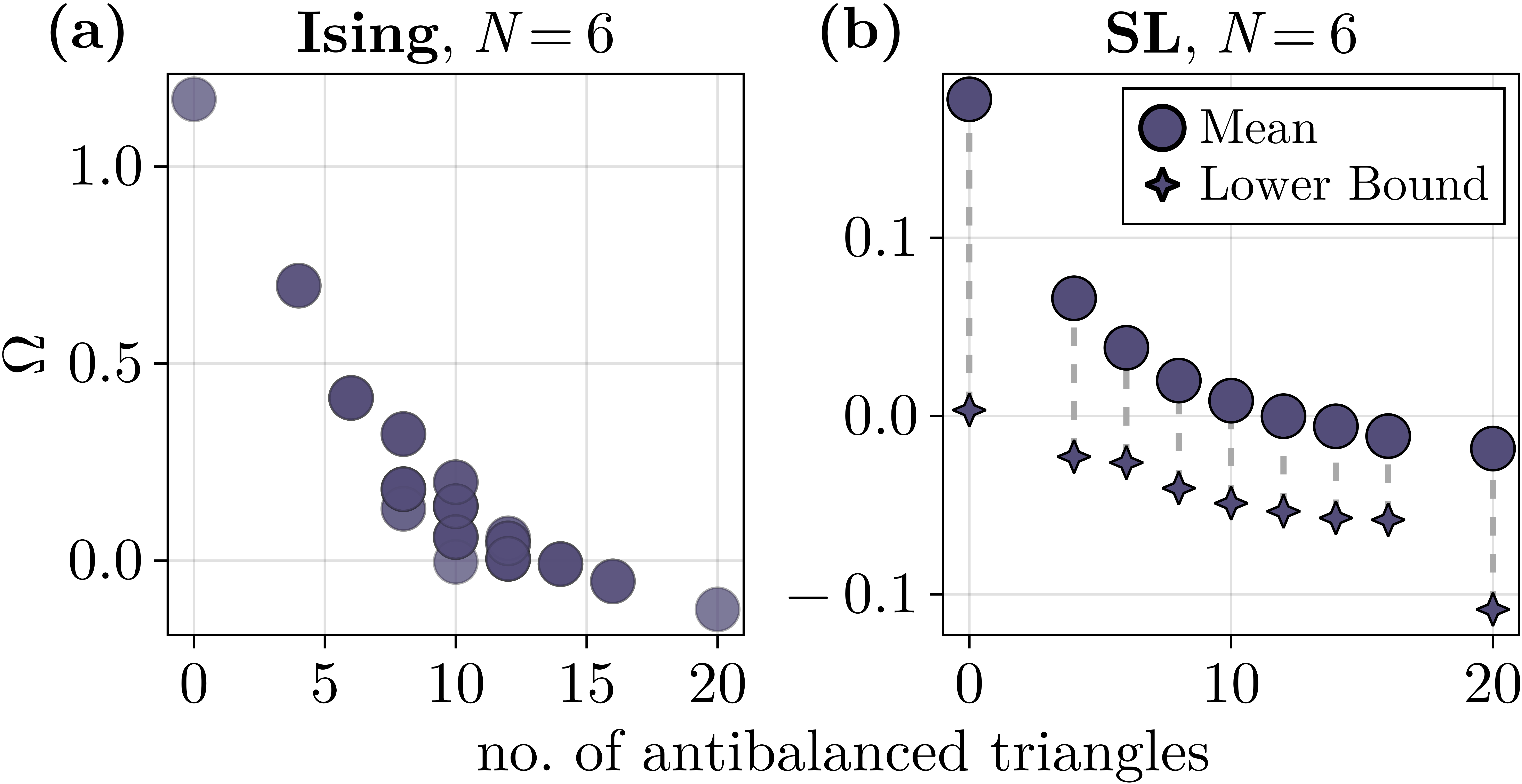}
    \caption{\textbf{Antibalanced interaction motifs characterize synergy-dominated complex systems with pairwise coupling.} Plots show $\Omega$ versus the number of antibalanced triangles in the pairwise coupling matrix. \textbf{(a)} For the Ising model we compute $\Omega$ exactly for each possible coupling configuration ($\beta=0.25$). Darker markers indicate multiple configurations with identical values of $\Omega$ and no. of antibalanced triangles \textbf{(b)} For the SL network of oscillators we report the mean $\Omega$ (circle) over $100$ initial conditions and the minimum $\Omega$ (star) found (lower bound) across configurations with a specific no. of antibalanced triangles. See SM~\cite{SM} for implementation details.}
    \label{fig:Ising-SL}
\end{figure}

\begin{figure}
    \centering
    \includegraphics[width=1\linewidth]{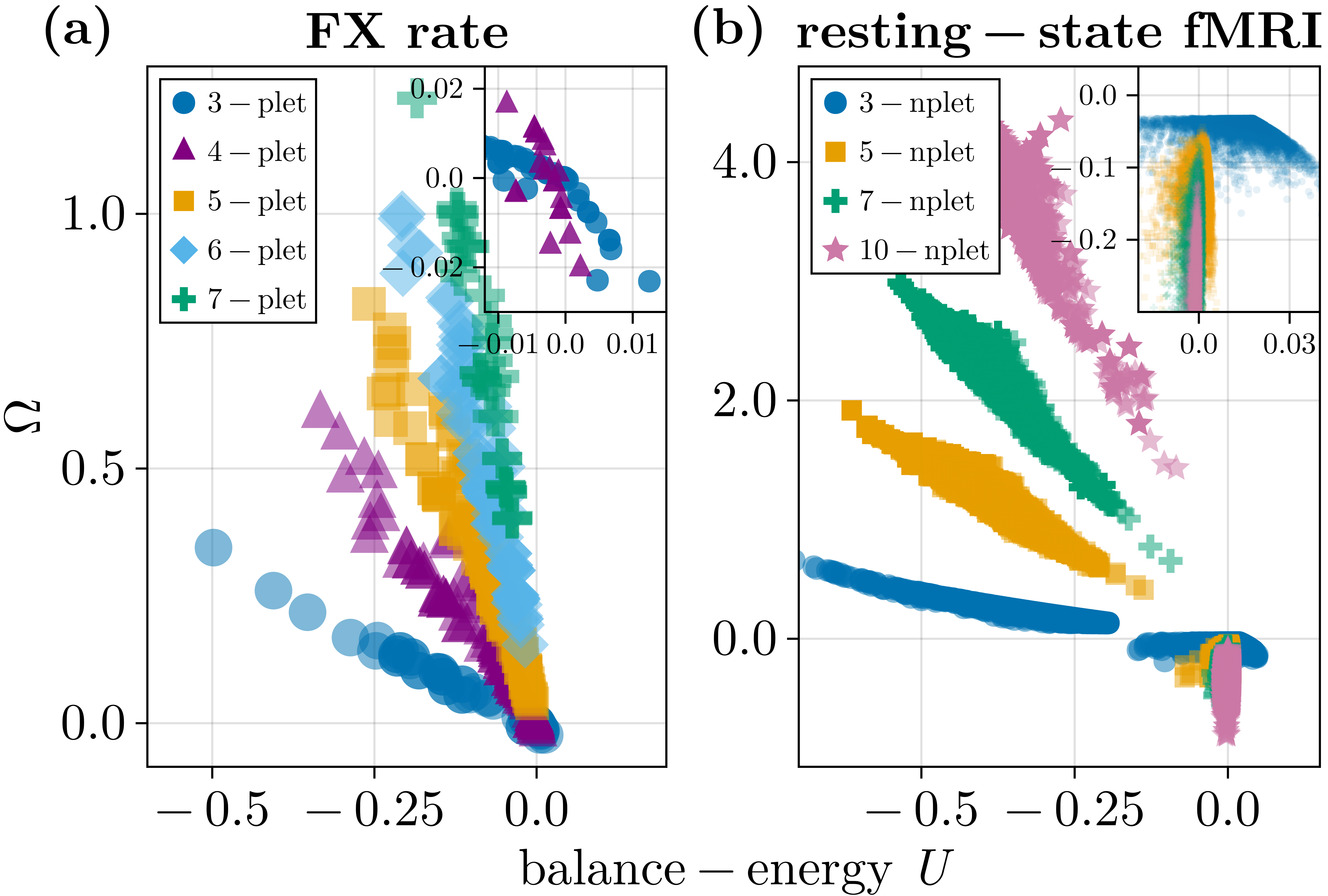}
    \caption{\textbf{Structural balance-energy predicts synergy: the most synergistic $N$-plets maximise $U$ and positive $U$ guarantees synergy-dominance.} \textbf{(a)} $N$-plets from a multivariate time-series of $9$ foreign-exchange (FX) rate logarithmic returns. \textbf{(b)} $N$-plets from resting-state fMRI functional-connectivity matrices ($229$ brain regions).}
    \label{fig:Real-World-Analysis}
\end{figure}

For both the Ising and SL models we performed a numerical exploration of all possible signed coupling structures, similar to that for the OU process above. For the Ising model, we studied $N$-spin systems without an external field. For $N<10$, all the relevant information theoretic quantities can be calculated exactly. For the SL networks, we numerically solved the system of equations as in~\cite{Cabral2022, Kringelbach2023, Luppi2024Competitive} (see also SM~\cite{SM}). Across the board, the coupling structures with the largest number of antibalanced triangles consistently result in those with the lowest O-information, in agreement with our predictions (Fig.~\ref{fig:Ising-SL}). Importantly, systems with zero antibalanced triangles consistently showed $\Omega>0$, confirming our analytical results placing antibalance as a necessary condition for synergy-dominance.

In the empirical datasets, where the underlying interaction matrix $A$ is not available, we studied the O-information as a function of the structural balance-energy $U$ of various $N$-plets (i.e., correlation matrix of subsystems) of size $N$ (Fig.~\ref{fig:Real-World-Analysis}). In both datasets, in which the majority of subsystems display $\rho(W)>1$ (indicative of strong interactions, see SM~\cite{SM}), our results demonstrate a systematic relationship between patterns of pairwise correlations (quantified by $U$) and high-order interdependencies (quantified by $\Omega$), confirming our analytical insights (Eq.~($4$) and Theorem $1$) that predominantly antibalanced subsystems with $U>0$ are synergy-dominated (and conversely, predominantly balanced systems with $U<0$ are redundancy-dominated). Specifically, results in Fig.~$4$ show that: ($i$) the most synergistic (redundant) subsystems have the highest (lowest) energy; ($ii$) if an $N$-plet has $U>0$, it is always synergy-dominated (Fig.~$4$ insets). To test the robustness of our conclusions, we additionally report in the SM~\cite{SM} a comparison between O-information estimates obtained using Gaussian copula and Kraskov estimators (which capture nonlinear dependencies), as well as further analyses on medical, epidemiological, financial, and physiological datasets.

\textit{Conclusion.}---We have established, for the first time, analytical sufficient and necessary conditions for the emergence of high-order behaviours from low-order mechanisms, providing a theoretical ground to a rapidly growing line of research~\cite{Battiston2021,Rosas2022,Milln2025, Murphy2025,Robiglio2025,Hahn2023,Faes2025,Decelle2025,Malizia2025}. Specifically, we proved that antibalanced correlations are sufficient for synergy-dominance in Gaussian systems and that antibalanced interaction motifs are necessary for synergy-dominance in OU processes. While our analytical results are derived in the weak interaction regime (except for Theorem $2$ which is valid for any $\rho(W)$), numerical simulations and empirical data analyses show that the same qualitative behaviour strongly persists beyond the weak interaction regime. Our results are therefore directly applicable whenever multivariate dynamics are modelled as approximately stationary linear processes with Gaussian statistics, which is a widely used approach in information-theoretic analyses of multivariate continuous data in neuroscience and biology. Moreover, our analyses of Ising and SL oscillator networks, together with the comparison between Gaussian and nonlinear Kraskov estimators on empirical datasets, indicate that the same relationship between antibalanced motifs and synergy extends beyond linear Gaussian models. These results provide a novel mechanistic link between low-order mechanisms and high-order interdependencies, establishing SBT as a natural framework for modelling and analysing synergistic complex systems with pairwise interactions. The expansion of the O-information in terms of signed closed walks has immediate implications for empirical analyses: ($i$) We expect our motifs analysis to point to faster and more principled algorithms for detecting synergistic subsystems in real datasets whenever the Gaussian approximation is justified \cite{belloli2025thoi, Varley2023, Puxeddu2025}. ($ii$) The leading, normalised term of the expansion is mathematically identical to the structural balance-energy $U$, explaining the similarities and uncovering novel connections between two previously disconnected lines of work, namely information-theoretic~\cite{Gatica2021, Varley2023C, Puxeddu2025} and SBT-based studies~\cite{Moradimanesh2021, Saberi2021, Talesh2023, Saberi2024}. We demonstrate this connection explicitly in the SM~\cite{SM} by reanalysing an existing SBT-based study using the O-information instead of $U$. Finally, our Ising results motivate exploring frustrated quantum magnets, including spin liquids~\cite{Norman2016, Broholm2020}, using quantum measures of high-order interdependencies~\cite{vanEnk2023, Javarone2024}.

\textit{Data availability}---The fMRI correlation matrices data which support the findings of this article is openly available~\cite{Rieck2021, RieckOSF}, as well as the multivariate time series financial data~\cite{CliffDataset} (see also~\cite{Cliff2023}). To construct the non-isomorphic signed graphs used throughout the study we used the openly available dataset~\cite{mckay_dataset}.
The code used to collect the data and produce the figures is available at~\cite{github_repo_synergistic_motifs}.

\putbib
\end{bibunit}

\clearpage




\renewcommand{\bibnumfmt}[1]{[S#1]}
\renewcommand{\citenumfont}[1]{S#1}

\newcommand{\sumeven}{\sum_{\substack{k=4 \\ \text{even }k}}}
\newcommand{\sumodd}{\sum_{\substack{k=5 \\ \text{odd }k}}}


\renewcommand{\thefigure}{S\arabic{figure}}
\setcounter{equation}{0}
\renewcommand{\theequation}{S\arabic{equation}}
\setcounter{secnumdepth}{3}
\renewcommand{\thesection}{\Roman{section}}
\setcounter{table}{0}
\renewcommand{\thetable}{S\arabic{table}}
\pagenumbering{roman}
\setcounter{page}{1}


\begin{bibunit}

\onecolumngrid

\begin{center}
    \textbf{\large Supplemental Material}
    \vspace{0.5cm}
\end{center}

Here, we provide a step-by-step derivation of all the information theoretic quantities (Sec.~\ref{sec:step-ny-step-static-gaussian-IT} and \ref{sec:step-ny-step-dynamical-gaussian-IT}), and details about the models and datasets used for the experimental validations (Sec.~\ref{sec:SM-OU-process} and \ref{sec:Experimental-Validation}). The code used to collect the data and produce the figures reported in the main text and here is available at~\cite{github_repo_synergistic_motifs}.

\section{Derivation of Static Gaussian systems analytical results}\label{sec:step-ny-step-static-gaussian-IT}
\subsection{Gaussian O-information derivation}\label{sec:gaussian-O-info}
Let $\bX=\{X_1,X_2,\dots,X_N\}$ be a multivariate Gaussian process of size $N$ with correlation matrix $\Sigma$, and let $\bX_{-i} = \bX\setminus{X_i}$ denote the same system with the $i$\textsuperscript{th} variable removed. The differential entropy of a Gaussian system can be written as~\cite{Cover2005}
\begin{equation}
    H(\bX)=\frac12\log\Big[(2\pi e)^N\det(\Sigma)\Big].
\end{equation}
Using this expression, we write the total correlation ($TC$) as
\begin{align*}
    TC(\bX) &= \sum_j^N\Big[H(X_j)\Big]-H(\bX) \\
    &= \frac12\log\big[(2\pi e)^N\big] - \frac12\log\Big[(2\pi e)^N\det(\Sigma)\Big] \\
    &= - \frac12\log\Big[\det(\Sigma)\Big].
\end{align*}
Then, the O-information can be written as,
\begin{align}
    \Omega(\bX) &= \sum_{i=1}^N TC(\bX_{-i}) - (N-2) TC(\bX), \label{eqn:o-info-only-tc} \\
    &= \sum_{i=1}^N \Big\{-\frac12\log\Big[\det(\Sigma_{-i})\Big]\Big\} - (N-2)\Big\{-\frac12\log\Big[\det(\Sigma)\Big]\Big\} \notag \\
    &= \frac{N-2}{2}\log\Big[\det(\Sigma)\Big] - \sum_{i=1}^N\frac12\log\Big[\det(\Sigma_{-i})\Big] \label{eqn:o-info-gaussian}
\end{align}
as reported in Eq.~($2$) in the main text.

\subsection{Proof of trace identities}\label{sec:appendix-trace-identity}
To isolate the leading term in Eq.~\eqref{eqn:o-info-gaussian}, we used the trace identity $\sum_i \tr[W_{-i}^k] = (N-k)\tr[W^k]$ valid for $k=2,3$. To derive these equations, recall that, given a matrix $M\in\mathbb{R}^{N\times{N}}$ with zeros along the diagonal, we have  $(M^k)_{ii}=\sum_{j_1,j_2,\dots,j_k-1}M_{i\,j_1}M_{j_1\,j_2}\dots M_{j_{k-1}\,i}$. Then we can write the trace of a subsystem $W_{-i}^2$ as
\begin{align*}
    \tr[W_{-i}^2] &= \tr[W^2] - \sum_{j\neq i} W_{ij}W_{ji} - \sum_{j\neq i} W_{ji}W_{ij} \\
    &= \tr[W^2] - 2\sum_{j\neq i} W_{ij}W_{ji},
\end{align*}
since $W$ is symmetric and $\mathrm{tr}[W^1]=0$. Similarly,
\begin{align*}
    \tr[W_{-i}^3] &= \tr[W^3] - \sum_{\substack{j,k \\ i\neq j\neq k}} W_{ij}W_{jk}W_{ki}  - \sum_{\substack{j,k \\ i\neq j\neq k}} W_{ji}W_{ik}W_{kj} - \sum_{\substack{j,k \\ i\neq j\neq k}} W_{jk}W_{ki}W_{ij} \\
    &= \tr[W^3] - 3\sum_{\substack{j,k \\ i\neq j\neq k}} W_{ij}W_{jk}W_{ki}.
\end{align*}
So that when summing over $i$ we obtain $\sum_i \tr[W_{-i}^2] = (N-2)\tr[W^2]$ and $\sum_i \tr[W_{-i}^3] = (N-3)\tr[W^3]$. While it is tempting to generalize $\sum_i \tr[W_{-i}^k] = (N-k)\tr[W^k]$ to any $k$, this identity already breaks down for $k=4$, since for $k\geq4$ closed walks on the graph with adjacency matrix $W$ can have repeated vertices. Thus, we introduce a general trace identity valid for any $k$ (Proposition 1).

\textit{Proposition 1}---For $k\in\mathbb{N}$, $k>0$, we have
\begin{align*}
    \sum_i\tr[W_{-i}^k] = N\tr[W^k] - \sum_{w\in\boldsymbol{w}^k}|w|\sigma(w),
\end{align*}
where $\boldsymbol{w}_i^k$ is the set of closed walks of length $k$ that include element $i$ at least once and $|w|$ is the number of unique nodes encountered along the closed walk $w$.

\textit{Proof of Proposition 1}---We begin by writing the trace of $W_{-i}^k$ (the submatrix $W$ with row and column $i$ removed) in terms of closed walks. The total sum of the weights of all closed walks of length $k$ in the full graph $G$ is
\begin{align*}
    \tr[W^k] &= \sum_i\,(W^k)_{ii} = \sum_{w\in\boldsymbol{w}^k}\sigma(w)
\end{align*}
If vertex $i$ is removed, the trace $\tr[W_{-i}^k]$ corresponds to the sum of weights of all closed walks of length $k$ that \emph{do not} visit vertex $i$ at any point. Therefore, $\tr[W_{-i}^k]$ can be written as the total sum of all walks' weights minus the sum of the weights of the walks that \emph{do} include vertex $i$:
\begin{equation}
    \tr[W_{-i}^k] = \sum_{w\in\boldsymbol{w}^k}\sigma(w) - \sum_{w\in\boldsymbol{w}_i^k}\sigma(w),\label{eqn:trace-element-removed}
\end{equation}
where $\boldsymbol{w}_i^k$ is the set of all closed walks of length $k$ that include element $i$ at least once. Next, we take the sum over all elements $i\in\{1, ,\dots,N\}$:
\begin{align*}
    \sum_{i=1}^N \tr[W_{-i}^k] &= \sum_{i=1}^N \left( \sum_{w\in\boldsymbol{w}^k}\sigma(w) - \sum_{w\in\boldsymbol{w}_i^k}\sigma(w) \right) \\
    &= N \sum_{w\in\boldsymbol{w}^k}\sigma(w) - \sum_{i=1}^N \sum_{w\in\boldsymbol{w}_i^k}\sigma(w)
\end{align*}
For the first term on the RHS, we know that $\sum_{w\in\boldsymbol{w}^k}\sigma(w) = \tr[W^k]$. For the second term, $\sum_{i=1}^N \sum_{w\in\boldsymbol{w}_i^k}\sigma(w)$, consider any closed walk $w \in \boldsymbol{w}^k$. This walk $w$ visits $|w|$ distinct nodes in general (only for $k=2,3$ it always visits exactly $2$ and $3$ nodes, respectively). For example, $w=\{1,2,1,2,1\}\in\boldsymbol{w}^4$ has $|w|=2$. When we sum over all $i$, this walk $w$ will be included in the inner sum $\sum_{w\in\boldsymbol{w}_i^k}\sigma(w)$ exactly $|w|$ times (once for each distinct node it visits). Therefore, the sum $\sum_{i=1}^N \sum_{w\in\boldsymbol{w}_i^k}\sigma(w)$ is equivalent to $\sum_{w\in\boldsymbol{w}^k}|w|\sigma(w)$. Substituting these expressions back into the equation, we obtain:
\begin{align}
    \sum_i\tr[W_{-i}^k] &= N\sum_{w\in\boldsymbol{w}^k}\sigma(w) - \sum_{w\in\boldsymbol{w}^k}|w|\sigma(w) \notag \\
    &= N\tr[W^k] - \sum_{w\in\boldsymbol{w}^k}|w|\sigma(w),
    \label{eqn:sum-over-i-traces-with-i-removed}
\end{align}
which completes the proof.\hfill\ensuremath{\blacksquare}

\subsection{Step-by-step proof of Theorem 1}\label{sec:link-IT-SBT-proof-lemma}
The key intuition to prove Theorem 1 is that all the terms $\tr[W^k]$ can be written in terms of closed walks of length $k$ (using Proposition 1). Then, in terms of closed walks, the difference between the full system $W^k$ and the subsystem $W_{-i}^k$ is simply whether a walk passes through node $i$, allowing us to simplify our expressions and identify the closed walks that contribute negatively in the expression for $\Omega$.

\textit{Theorem 1}---If $G$ is antibalanced, then $\Omega(\bX)<0$.

\textit{Proof of Theorem 1 (step-by-step)}---We begin by substituting the log determinant expansion shown in Eq.~\eqref{eqn:o-info-only-tc} into Eq.~\eqref{eqn:o-info-gaussian} to obtain
\begin{align}
    \Omega(\bX) &= \frac{N-2}{2}\Bigg[- \frac{\mathrm{tr}[W^2]}{2} + \frac{\mathrm{tr}[W^3]}{3} - \frac{\mathrm{tr}[W^4]}{4} + \frac{\mathrm{tr}[W^5]}{5} \dots  \Bigg] \notag \\
    &\quad + \frac12\Bigg[-\frac{\sum_i\mathrm{tr}[W_{-i}^2]}{2}+\frac{\sum_i\mathrm{tr}[W_{-i}^3]}{3}+\frac{\sum_i\mathrm{tr}[W_{-i}^4]}{4} - \frac{\sum_i\mathrm{tr}[W_{-i}^5]}{5} +  \dots \Bigg], \label{eqn:O-info-expansion-no-simplification}
\end{align}
Before using Proposition 1 and substitute Eq.~\eqref{eqn:sum-over-i-traces-with-i-removed} into Eq.~\eqref{eqn:O-info-expansion-no-simplification} note that, in general, if $k$ is even each closed walk has at least $2$ distinct vertices, while if $k$ is odd each closed walk has at least $3$ distinct vertices. Then, we can rewrite Eq.~\eqref{eqn:sum-over-i-traces-with-i-removed} as
\begin{align}
    \sum_i\tr[W_{-i}^k] = (N-2)\tr[W^k] - \sum_{w\in\boldsymbol{w}^k}(|w| - 2)\sigma(w) \label{eqn:sum-over-i-traces-with-i-removed-v2}
\end{align}
since each closed walk $w$ has at least $2$ distinct vertices for even $k$ and at least $3$ distinct vertices for odd $k$ (recall also $\tr[W^k]=\sum_{w\in\boldsymbol{w}^k}\sigma(w)$). Then, we insert Eq.~\eqref{eqn:sum-over-i-traces-with-i-removed-v2} into Eq.~\eqref{eqn:O-info-expansion-no-simplification} to obtain
\begin{align*}
    \Omega(\bX) = \frac16\tr[W^3] &+ \frac12\Bigg\{-\frac14\sum_{w\in\boldsymbol{w}^4}(|w|-2)\sigma(w) -\frac16\sum_{w\in\boldsymbol{w}^6}(|w|-2)\sigma(w)- \dots\Bigg\} \\
    &+ \frac12\Bigg\{+ \frac15\sum_{w\in\boldsymbol{w}^5}(|w|-2)\sigma(w) + \frac17\sum_{w\in\boldsymbol{w}^7}(|w|-2)\sigma(w) + \dots\Bigg\},
\end{align*}
where all the terms of the form $(N-2)\tr[W^k]$ cancelled out. This expression can be more concisely written as
\begin{align}
    \Omega(\bX) &= \frac16\tr[W^3] + \frac12\sum_{k=4}^\infty\frac{1}{k}\Big[(-1)^{k-1}\sum_{w\in\boldsymbol{w}^k}(|w|-2)\sigma(w)\Big], \notag \\
    &= \sum_{k=3}^\infty\Bigg[\frac{(-1)^{k-1}}{2k}\sum_{w\in\boldsymbol{w}^k}(|w|-2)\sigma(w)\Bigg].\label{eqn:o-info-full-walk-expansion}
\end{align}
Now, recall that if $G$ is antibalanced, then every closed walk $w\in\boldsymbol{w}^k$ in $G$ satisfies $(-1)^{k-1}\sigma(w)\leq0$. Thus, if $G$ is antibalanced, then $\Omega \leq 0$, completing the proof. \hfill\ensuremath{\blacksquare}

\begin{figure}[h]
    \centering
    \includegraphics[width=0.6\linewidth]{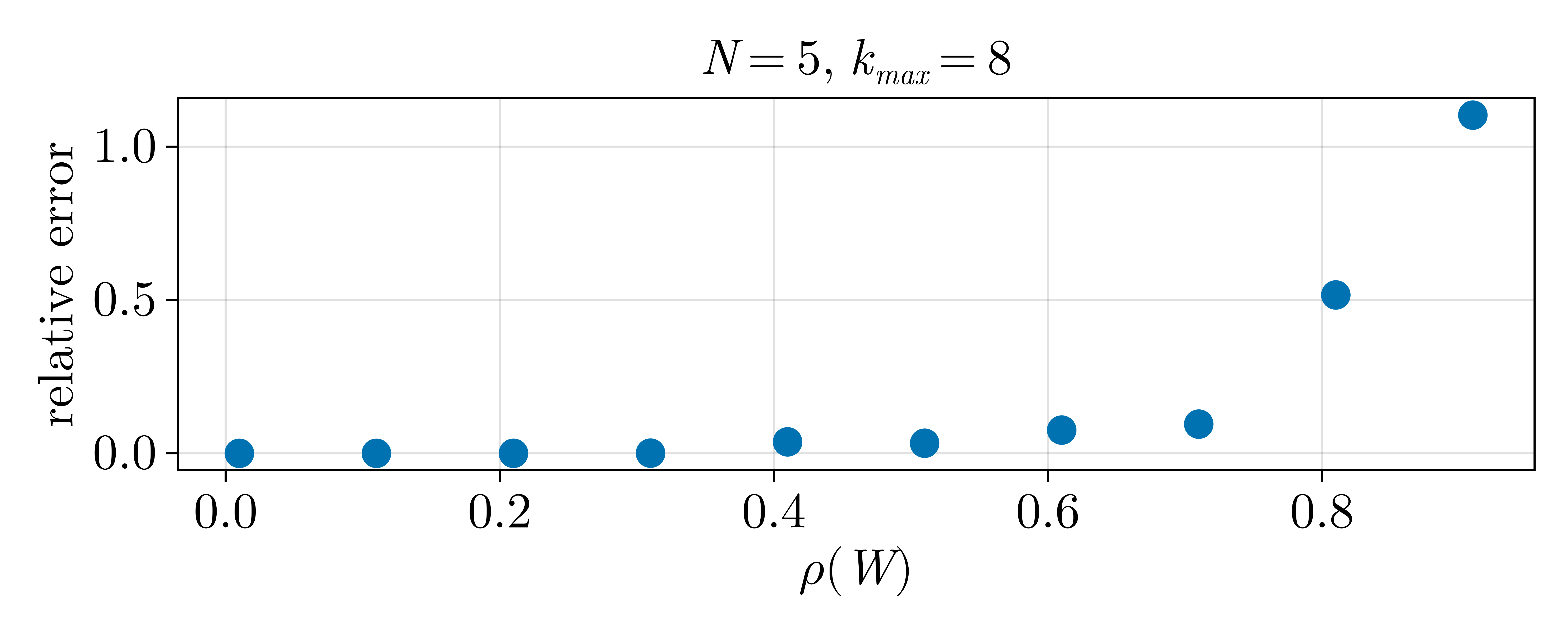}
    \caption{Mean relative error between $\Omega$ computed using Eq.~($1$) in the main text and Eq.~\eqref{eqn:o-info-full-walk-expansion} up to walks of length $k=k_{max}=8$ for different spectral radii $\rho(W)$.}
    \label{fig:o-info-final-walk-approximation}
\end{figure}

In Fig.~\ref{fig:o-info-final-walk-approximation} we show how Eq.~\eqref{eqn:o-info-full-walk-expansion} approximates the O-information.
Additionally, for the subsequent section, it useful to isolate the cubic term in Eq.~\eqref{eqn:o-info-full-walk-expansion}, and split the summation over all $k$ as
\begin{align}
    \Omega(\bX) &= \frac12 \Bigg[\frac{1}{3}\sum_{w\in\boldsymbol{w}^3}\sigma(w) - \sumeven \frac{1}{k}\sum_{w\in\boldsymbol{w}^k}(|w|-2)\sigma(w) + \sumodd \frac{1}{k}\sum_{w\in\boldsymbol{w}^k}(|w|-2)\sigma(w) \Bigg].\label{eqn:O-info-walk-expansion-splitted}
\end{align}
In this expression, the lowest order term is the same as the structural balance-energy metric $U$ (opposite sign and up to a normalization constant)
\begin{equation}
    U= -\frac{1}{\binom{N}{3}}\sum_{\{i,j,k\}}{W_{ij}W_{jk}W_{ik}},\label{eqn:structural-energy}
\end{equation}
used in e.g.,~\cite{Marvel2009, Moradimanesh2021, Masoomy2021, Talesh2023, Saberi2021, Saberi2024}. Particularly, Eq.~\eqref{eqn:O-info-walk-expansion-splitted} highlights the closed walks that contribute synergistically: ($i$) even-length closed walks with $\sigma(w)>0$, ($ii$) odd-length closed walks with $\sigma(w)<0$. In Sec.~\ref{sec:R-bounds} we study the upper bound of the higher-order terms $R_4(W)=\sumodd \frac{1}{k}\sum_{w\in\boldsymbol{w}^k}(|w|-2)\sigma(w) - \sumeven \frac{1}{k}\sum_{w\in\boldsymbol{w}^k}(|w|-2)\sigma(w)$ in Eq.~\eqref{eqn:O-info-walk-expansion-splitted}.

\section{Derivation of Dynamical Gaussian systems analytical results} \label{sec:step-ny-step-dynamical-gaussian-IT}
In this section, we provide a step-by-step proof of Theorem 2, the analytical bounds on the higher-order terms $R_4$ in the expansion of $\Omega$, and numerical tests to identify the range of $\rho(W)$ for which the cubic terms dominate in the expansion of $\Omega$ (Eq.~\eqref{eqn:o-info-full-walk-expansion}).

\subsection{Step-by-step proof of Theorem 2}\label{sec:dynamical-systems-N-3}

\textit{Theorem 2}---Let $\bX$ be a stable OU process with $A\in\mathbb{R}^{3\times3}$. If $\Omega(\bX)<0$, then $A$ is antibalanced.

\textit{Proof of Theorem 2}---Let $\mathrm{corr}(X) = \mathrm{diag}(X)^{-1/2}\;X\;\mathrm{diag}(X)^{-1/2}$. We start by showing that, for any interaction matrix $A$ with self-interactions, i.e., $A_{ii}\neq0$, there is a matrix $B$ with $B_{ii}=0$ and $B_{ij}=A_{ij}/\sqrt{(1-A_{ii})(1-A_{jj})}$ such that the stationary correlation of the OU process $\Sigma=\mathrm{corr}((\mathbb{I}-A)^{-1})=\mathrm{corr}((\mathbb{I}-B)^{-1})$.

Let us define the matrix $\Xi$ with elements $\Xi_{ii}=\sqrt{1-A_{ii}}$ such that $(\mathbb{I}-A)=\Xi\;(\mathbb{I}-B)\;\Xi$. For any positive diagonal matrix $D$ we have
\begin{align*}
    \mathrm{corr}(DXD) &= \mathrm{diag}(DXD)^{-1/2}\;DXD\;\mathrm{diag}(DXD)^{-1/2} \\
    &= D^{-1/2}\;\mathrm{diag}(X)^{-1/2}\;D^{-1/2}\;DXD\;D^{-1/2}\;\mathrm{diag}(X)^{-1/2}\;D^{-1/2} \\
    &= \mathrm{diag}(X)^{-1/2}D^{-1}DXDD^{-1}\mathrm{diag}(X)^{-1/2} \\
    &= \mathrm{corr}(X),
\end{align*}
since the diagonal matrices commute and $D^{-1}D=DD^{-1}=\mathbb{I}$. Then we have
\begin{align*}
    \Sigma &= \mathrm{corr}((\mathbb{I}-A)^{-1}) \\
    &= \mathrm{corr}(\Xi^{-1}(\mathbb{I}-B)^{-1}\Xi^{-1}) \\
    &= \mathrm{corr}((\mathbb{I}-B)^{-1}),
\end{align*}
such that $\Omega(A)=\Omega(B)$ since $\mathrm{sign}(A_{ij})=\mathrm{sign}(B_{ij})$. Next, we prove the theorem for the case with no self-interactions i.e., $A_{ii}=0$. Let us write $(\mathbb{I}-A)^{-1}=C\;(\det(\mathbb{I}-A))^{-1}$, where $C=\mathrm{adj}(\mathbb{I}-A)^{\mathsf{T}}$. Then, each element of $\Sigma$ can be written as
\begin{equation}
    \Sigma_{ij} = 
\begin{cases}
    \frac{C_{ij}}{\sqrt{C_{ii}C_{jj}}} & \text{if } i \neq j \\
    1 & \text{if } i = j 
\end{cases} \label{eqn:sigma-elements-OU-process_N-3}
\end{equation}
where for every pair $i,j\in\{1,2,3\},\;i\neq{j},$ and $k = \{1,2,3\}\setminus\{i,j\}$ we have $C_{ii} = 1-A_{jk}^2> 0$, and $C_{ij} = A_{ij} + A_{ik}A_{jk}$.
Recall the synergistic condition for a static Gaussian system of size $N=3$. Using $a=\Sigma_{12}, b=\Sigma_{13},$ and $c=\Sigma_{23}$, $\Omega<0$ if
\begin{equation*}
    2abc < (ab)^2 + (ac)^2 + (bc)^2 - (abc)^2.
\end{equation*}
Let $x=A_{12},\,y=A_{13},\,z=A_{23}$. After substituting the elements from Eq.~(\ref{eqn:sigma-elements-OU-process_N-3}) into the synergistic condition, the LHS becomes
\begin{align*}
    2abc = 2\frac{(x+yz)(y+xz)(z+xy)}{(1-x^2)(1-y^2)(1-z^2)},
\end{align*}
while the RHS becomes
\begin{align*}
    (ab)^2+(ac)^2+(bc)^2-(abc)^2 &= \frac{(x+yz)^2(y+xz)^2}{(1-x^2)(1-y^2)(1-z^2)^2} + \frac{(x+yz)^2(z+xy)^2}{(1-x^2)(1-y^2)^2(1-z^2)} \\
    & \quad + \frac{(y+xz)^2(z+xy)^2}{(1-x^2)^2(1-y^2)(1-z^2)} + \frac{(x+yz)^2(y+xz)^2(z+xy)^2}{(1-x^2)^2(1-y^2)^2(1-z^2)^2}\,.  \\
\end{align*}
To simplify these, we multiply both sides by the common denominator $(1-x^2)^2(1-y^2)^2(1-z^2)^2$ which is always positive and non-zero for Schur-stable $A$. Taking the difference between the RHS and LHS we obtain
\begin{align*}
    &(1 - x^2) (1 - y^2) (x + y z)^2 (y + x z)^2 + (1 - x^2) (1 - z^2) (x + y z)^2 (z + x y)^2 + (1 - y^2) (1 - z^2) (z + x y)^2 (y + x z)^2 \\
    &\quad\quad - (z + x y)^2 (y + x z)^2 (x + y z)^2 - 2 (z + x y) (y + x z) (x + y z) (1 - x^2) (1 - y^2) (1 - z^2) > 0 \\
\end{align*}
This expression can be written in the form $k_1^2k_2>0$, where
\begin{align*}
    k_1 &= (-1 + x^2 + y^2 + 2 x y z + z^2), \notag \\
    k_2 &= (x y z)^2 - (x y)^2 - (x z)^2 - (y z)^2 - 2 x y z. \notag
\end{align*}
Since $k_1^2$ is always positive, we require $k_2$ to be positive for the condition to be satisfied, or equivalently
\begin{equation*}
     - 2 x y z > + (x y)^2 + (x z)^2 + (y z)^2 - (x y z)^2.
\end{equation*}
For Schur-stable A, the RHS is always positive, so this expression is satisfied only if $xyz<0$. Thus, we conclude that if $\Omega(\bX)<0$ then $A$ is antibalanced. The same conclusion holds for the case with self-interactions, i.e., $A_{ii}\neq0$, since $\Sigma=\mathrm{corr}((\mathbb{I}-A)^{-1})=\mathrm{corr}((\mathbb{I}-B)^{-1})$ where $B_{ii}=0$.\hfill\ensuremath{\blacksquare}

\subsection{Sufficiency condition for cubic dominance}\label{sec:R-bounds}
In this section, we study the upper bound of the higher-order terms $|R_4(W)|$ in the expansion of $\Omega$ (Eq.~\eqref{eqn:O-info-walk-expansion-splitted} and Eq.~($4$) in the main text) for the case of balanced correlation matrices $\Sigma$. Specifically, we provide two sufficient conditions for cubic dominance, i.e, when $|\frac16\tr[W^3]| > |R_4(W)|$: a practical, checkable inequality for finite interaction strength and an asymptotic bound as $\rho(W)\to0$.

\subsubsection{Notation}
In what follows, we use the short notation $S_{\text{even}}$ to denote the sum over even $k$ terms in Eq.~\eqref{eqn:O-info-walk-expansion-splitted}, and $S_{\text{odd}}$ to denote the sum over odd $k>3$ terms, i.e.,
\begin{align*}
    S_{\text{even}} &= \sum_{\substack{k=4 \\ \text{even }k}} \frac{1}{2k}\sum_{w\in\boldsymbol{w}^k}(|w|-2)\sigma(w) \\
    S_{\text{odd}} &= \sum_{\substack{k=5 \\ \text{odd }k}} \frac{1}{2k}\sum_{w\in\boldsymbol{w}^k}(|w|-2)\sigma(w).
\end{align*}
Then, Eq.~\eqref{eqn:O-info-walk-expansion-splitted} can be written as
\begin{align}
    \Omega(\bX) &= \frac16\tr[W^3] + R_4(W) \label{eqn:o-info-cubic-term-plus-R}\\
    \text{where}&\quad R_4(W) = S_{\text{odd}} - S_{\text{even}}.\label{eqn:R-even-minus-odd}
\end{align}
In general, each term in $S_{\text{odd}}$ and $S_{\text{even}}$ can either be negative or positive due to the signed $\sigma(w)$ contributions. However, for the case of balanced $G$, for each closed walk $w$ we have $\sigma(w)>0$ such that $S_{\text{odd}}, S_{\text{even}} > 0$, allowing us to simply our expressions and find precise bounds for the even and odd series.

\subsubsection{Practical bound}

First, we obtain the lower bounds for $S_{\text{even}}$ and $S_{\text{odd}}$.

Recall that $2\leq|w|\leq k$ if $w\in\boldsymbol{w}^k$ and $k$ is even, while if $k$ is odd we have $3\leq|w|\leq k$. Thus, for balanced $G$, we obtain the lower bound by setting $|w|=3$ for $S_{\text{odd}}$ and $|w|=2$ for $S_{\text{even}}$, such that
\begin{align*}
    S_{\text{odd}} &\geq \sumodd\frac{1}{2k}\sum_j\lambda_j^k = L_{\mathrm{odd}} \\
    S_{\text{even}} &\geq 0,
\end{align*}
where we used $\sum_{w\in\boldsymbol{w}^k}\sigma(w)=\tr[W^k]=\sum_j\lambda_j^k$.

Secondly, we obtain the upper bounds for $S_{\text{even}}$ and $S_{\text{odd}}$ by setting $|w|=k$ for both $S_{\text{odd}}$ and $S_{\text{even}}$, obtaining
\begin{align}
    S_{\text{odd}} &\leq \sumodd\frac{k-2}{2k}\sum_j\lambda_j^k = U_{\mathrm{odd}} \label{eqn:upperbound-step-1} \\
    S_{\text{even}} &\leq \sumeven\frac{k-2}{2k}\sum_j\lambda_j^k = U_{\mathrm{even}}. \label{eqn:upperbound-step-2}
\end{align}

\subsubsection{Sufficiency test and numerical result}
Using Eq.~\eqref{eqn:R-even-minus-odd}, we combine the upper and lower bounds for $S_{\text{odd}}$ and $S_{\text{even}}$, to obtain
\begin{align}
    |R_4(W)| \;&\leq\; \max\big\{\,U_{\mathrm{even}} - L_{\mathrm{odd}},\; U_{\mathrm{odd}}\,\big\}\label{eqn:upper-bound-R-1}
\end{align}
A \emph{practical sufficiency test} for cubic dominance is then
\begin{equation}
  \left|\tfrac16\,\tr[W^{3}]\right| \;>\; \max\big\{\,U_{\mathrm{even}} - L_{\mathrm{odd}}, \;U_{\mathrm{odd}}\,\big\}.
  \label{eqn:cubic-dominance-test}
\end{equation}
We can use this inequality to estimate the range of interaction strengths $\rho(A)$ in OU processes for which the cubic term in Eq.~\eqref{eqn:o-info-cubic-term-plus-R} dominates. To do this, for each $N$, we sample over $10^{6}$ balanced Schur-stable matrices $A$ of various interaction strength $\rho(A)$ (see Sec. \ref{sec:SM-OU-process}). For each of these, we compute the stationary covariance of the OU process, and calculate the mean relative difference
\begin{equation}
    D(\rho(A)) = \frac{|\frac16\mathrm{tr}[W^3]| - \max\{\,U_{\mathrm{odd}},\ U_{\mathrm{even}}-L_{\mathrm{odd}}\,\}}{|\frac16\mathrm{tr}[W^3]| + \max\{\,U_{\mathrm{odd}},\ U_{\mathrm{even}}-L_{\mathrm{odd}}\,\}}\in[-1,1].
\end{equation}
When $D>0$, the cubic term is guaranteed to be larger than the higher-order terms $R_4$. We show our numerical results in Fig.~\ref{fig:weak-interactions-regime}, where we plot the difference $D$ as a function of $\rho(A)$.  Our numerical analysis confirms that there is a wide range of $\rho(A)$ for which the cubic term dominates Eq.~\eqref{eqn:o-info-cubic-term-plus-R} for any $N$ we studied (i.e., when $D>0$).

\begin{figure}[h]
    \centering
    \includegraphics[width=\linewidth]{figures/SM/weak_interactions_regime.png}
    \caption{\textbf{Cubic term dominance (when $D>0$) for various OU processes of size $N$}. We show $D$ as a function of the interaction strength $\rho(A)$ for: (a) $N=4$; (b) $N=5$; (c) $N=6$; (d) $N=7$; (e) $N=8$.}
    \label{fig:weak-interactions-regime}
\end{figure}

\subsubsection{Asymptotic bound}
Let $\rho(W)=\max_j\{|\lambda_j|\}$ and note that $|\tr[W^k]|\leq N\rho(W)^k$. Then, from Eq.~\eqref{eqn:upper-bound-R-1}, we have
\begin{align*}
    |R_4(W)| &\leq \mathcal{O}(\rho(W)^4)\quad\text{as}\quad\rho(W)\to0.
\end{align*}

\section{Ornstein–Uhlenbeck process implementation and numerical results}\label{sec:SM-OU-process}
\subsection{Methodology}\label{sec:SM-OU-process-methodology}
To test the relationship between the O-information and antibalanced structures in the interaction matrix, we computed $\Omega$ using the closed-form stationary covariance $\tilde\Sigma$ of the OU process with interaction matrix $A$ (all-to-all couplings) for all possible sign patterns in $A$. More specifically:
\begin{enumerate}
    \item we constructed the set $\boldsymbol{G}_N$ of all possible non-isomorphic complete signed graphs of size $N$ (see details below),
    \item for each of these graphs $G_i\in\boldsymbol{G}_N$, we sampled $1000$ Schur stable matrices with the particular sign patterns in the off-diagonals given by the adjacency matrix of the graph $G_i$ (see details below),
    \item finally, we computed the O-information using Eq.~\eqref{eqn:o-info-only-tc} and the correlation matrix obtained for each $A$.
\end{enumerate}

Note, as highlighted in the main text, in this work we restrict our analysis to \emph{complete} signed interaction networks (all-to-all couplings). This choice allowed us to explore the full range of possible signed patters: from the most balanced to the most antibalanced.

It should be noted that there are many ways to sample Schur stable interaction matrices $A$. Here, we show in Fig.~\ref{fig:OU-process-additional-figures} additional results for interaction matrices with different average spectral radii $\rho(A)$ and with or without self-interactions. Importantly, as shown in Fig.~\ref{fig:OU-process-additional-figures}, our results remain unaffected. For different interaction strengths, the main qualitative differences we observe are ($i$) the behaviour of the lower bound and ($ii$) the maximum mean value of $\Omega$. In the strong–interaction regime, approximately defined here as $\rho(A)>0.5$ (see Fig.~S4) when the number of antibalanced triangles is small the system displays much higher values of $\Omega$ than for the weak interactions case. As the number of antibalanced triangles increases, the lower bound rapidly approaches a minimum; this effect can be clearly observed for $N>6$ and $\rho(A)>0.5$. These nonlinear behaviours can be explained by the dominance of nonlinear contributions in the strong interactions regime, where $|\frac{1}{6}\mathrm{tr}[W^3]|>|R_4(W)|$ does not hold. In contrast, in the weak interactions regime (approximately $\rho(A)<0.5$) both the mean and the lower bound decrease approximately linearly as the number of antibalanced triangles increases (see the example for $N=8$ and $\rho(A)\approx0.37$). Additionally, we note that the slight increase of the empirical lower bound for large numbers of antibalanced triangles (for $N>6$) is a finite-sampling effect. Although our sampling algorithm guarantees Schur stability of the interaction matrices for any number of antibalanced triangles, it does not sample matrices uniformly conditional on a fixed number of frustrated triangles. Consequently, the configurations with the largest number of antibalanced triangles and lowest $\Omega$ can be under-represented, and the empirical minimum of $\Omega$ can be slightly higher than the true minimum.

\begin{figure}
    \centering
    \includegraphics[width=0.7\linewidth]{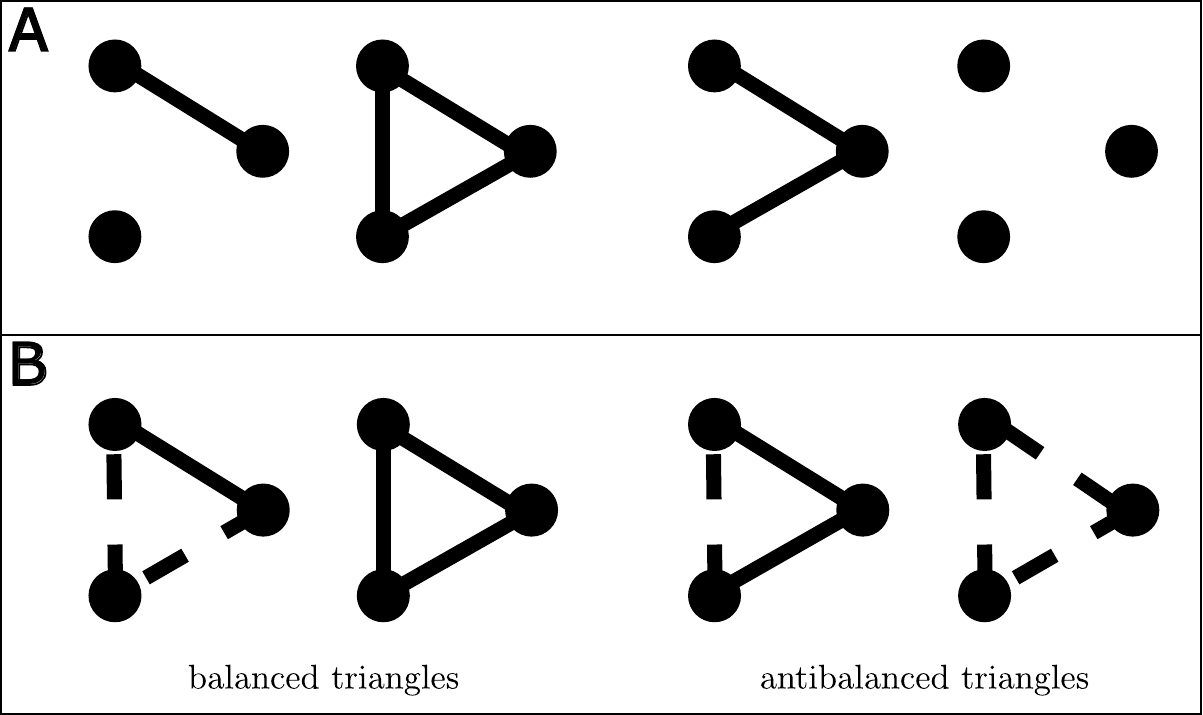}
    \caption{(\textbf{A}) All possible non-isomorphic graphs of size $N=3$ (unsigned, non-complete). (\textbf{B}) All possible complete signed graphs of size $N=3$. The product of the edges in the balanced triangles is positive, while the product of the edges in the antibalanced triangles is negative.)}
    \label{fig:triangle-configurations}
\end{figure}

\textit{Non-isomorphic complete signed graphs.} We constructed the set $\boldsymbol{G}_N$ using the non-isomorphic simple graphs dataset publicly available in~\cite{mckay_dataset}. This dataset contains the set of all non-isomorphic simple graphs of size $N<10$, which are unsigned and non-complete. Thus, we obtained the set of all non-isomorphic \emph{complete} signed graphs by substituting the missing edges with a negative edge (we illustrate this in Fig.~\ref{fig:triangle-configurations} for $N=3$).

\textit{Sampling Schur stable matrices}. Let $\mathbf{S}_N$ be the set of adjacency matrices of all possible non-isomorphic unweighted complete signed graphs in $\boldsymbol{G}_N$. To generate a Schur stable matrix $A$ with a desired sign pattern $S_i\in\mathbf{S}_N$, we can sample a random symmetric matrix $M$, multiply it element-wise by $S_i$, and then divide each element by the largest absolute eigenvalue to ensure stability, such that $A = (M \circ S_i)/\lambda_{\max}$ where $\circ$ denotes element-wise multiplication between matrices. This method ensures that the largest absolute eigenvalue of the resulting interaction matrix $A$ will be less than (but very close to) unity, i.e., the spectral radius of the resulting matrix $A$ is $\rho(A) < 1$. To study interaction matrices with smaller spectral radii we can add a parameter $\epsilon$, such that
\begin{equation}
    A = \frac{M \circ S_i}{\lambda_{\max}+\epsilon}, \label{eqn:interaction-matrix-formula}
\end{equation}
For Fig.~$2$ in the main text we use $\epsilon = 0.4$ to fix the spectral radius of the interaction matrices for $N=8$ to be $\rho(A) = 0.907 \pm 0.009$. To see how the spectral radius affects the O-information, see in Fig.~\ref{fig:OU-process-additional-figures} the case for $N=7$ with $\epsilon = 0.4$ and with $\epsilon = 0.001$.

Using this procedure, we collected data for $N\in\{4,5,6,7,8,9\}$, which we show in Fig.~\ref{fig:OU-process-additional-figures}. Note, two distinct structures $S_i$ and $S_j$ may have the same number of antibalanced triangles, yet be non-isomorphic. Let $m$ denote the number of antibalanced triangles and consider the case for $N=4$ (Fig.~\ref{fig:OU-process-additional-figures}) where $3$ structures have $m=0$, $3$ structures have $m=4$, and $5$ structures have $m=2$. Let $s(m)$ be the number of samples (i.e., the values of $\Omega$) collected for each $m$. Then for the case for $N=4$ we have $s(0) = 3000, s(4) = 3000$, and $s(2) = 5000$. To avoid sample size effects we only used $3000$ samples for each $m$ to compute the statistics we reported (similarly for the $N>4$ cases).

\begin{figure}
    \centering
    \includegraphics[width=1\linewidth]{figures/SM/OU_process_N_4_to_7.png}
    \caption{We show $\Omega$ as a function of the no. of antibalanced triangles in the interaction matrix of OU processes of size $N$. Each mean value and lower bound is computed over $10^5$ independent realizations of $A$ with a specific no. of antibalanced triangles. Lower bounds indicate the lowest $\Omega$ encountered in our numerical exploration across all configurations with a specific no. of antibalanced triangles. Each mean $\Omega$ is coloured according to the balance-energy value (Eq.~($5$)) of the corresponding interaction matrix $A$. Lower bounds are coloured with the energy value of the configuration that resulted in the lowest recorded Ω(X), not the highest recorded energy (not shown). (\textbf{Top row}) Results for $N=4,5,6$ using interaction matrices constructed with $\epsilon = 0.5$ (Eq.~\eqref{eqn:interaction-matrix-formula}), with mean interaction strengths corresponding to $\langle \rho(A) \rangle \approx 0.78,\,0.81,$ and $0.83$. (\textbf{Middle row}) Results for $N=7$ using $\epsilon = 0.5$ and $\epsilon = 0.001$, with $\langle \rho(A) \rangle \approx 0.84$ and $\approx 0.999$, respectively, and for the case without self-interactions ($A_{ii}=0\ \forall i$, with $\langle \rho(A) \rangle \approx 0.84$). (\textbf{Bottom row}) Results for $N=8,9$ using $\epsilon \in \{5,\,0.001,\,0.5\}$, corresponding to $\langle \rho(A) \rangle \approx 0.37,\,0.999,$ and $0.86$, respectively.}
    \label{fig:OU-process-additional-figures}
\end{figure}

\section{Experimental validation: model implementations and empirical datasets}\label{sec:Experimental-Validation}
Here, we describe more in details the synthetic and empirical datasets used to validate our analytical results. For both the Ising and SL models, we used all possible non-isomorphic complete signed graphs $\boldsymbol{G}_N$ to construct the coupling structures, as we did for the OU process numerical study (see Sec.~\ref{sec:SM-OU-process-methodology}).

\subsection{Ising Model}
\textit{Model definition.} We consider an Ising model of $N$ binary spins $s_i=\{-1,+1\}$ with no external field. The Hamiltonian is
\begin{equation}
    \mathcal{H} = \sum_{\langle i,j\rangle}J_{ij}s_is_j
\end{equation}
where the sum runs over all spin pairs $\langle i,j\rangle$. Here, we consider all-to-all coupling schemes only (as in the OU process study) and allow the off-diagonal terms $J_{ij}$ to be either positive or negative: $J_{ij}\in\{1,-1\}$. The Boltzmann weight of a micro-state $s=(s_1, s_2, \dots, s_N)$ at inverse temperature $\beta$ is $b(s)=e^{-\beta\mathcal{H}(s)}$. For $N\leq10$, the state space $S=\{-1,+1\}^N$ can be enumerated exhaustively, allowing us to compute the partition function $Z=\sum_{s\in{S}}b(s)$ and the probability of each state as $p(s) = b(s)/Z$.

\textit{Information theoretic quantities for the Ising model.} With the state space probability distribution, we can compute the entropy of the full system: $H=-\sum_{s}p(s)\log{p(s)}$. Similarly, we obtain the entropy of the subsystems $H_{-i}$ by evaluating $H_{-i}=-\sum_{s_{-i}}p_{-i}(s_{-i})\log{p_{-i}(s_{-i})}$ for each $i$, where $s_{-i}=(s_1, \dots, s_{i-1}, s_{i+1}, \dots, s_N)$ denotes a state with the element $i$ removed, whose probability is obtained by marginalising over spin $s_{i}$ i.e., $p_{-i}(s_{-i})=\sum_{s_i=\pm1}p(s_{-i},s_{i})$. With these expressions, we can compute the total correlation using $TC=\sum_{i}{H_i}-H=N-H$ (since the entropy of a spin is $H_i = 1$ in the absence of an external field), and the total correlation of the subsystems using $TC_{-i}=\sum_{j\neq{i}}{H_j}-H_{-i}=(N-1)-H_{-i}$. Then, the O-information is obtained via $\Omega=\sum_{i}TC_{-i}-(N-2)TC$.

\begin{figure}[h]
    \centering
    \includegraphics[width=.8\linewidth]{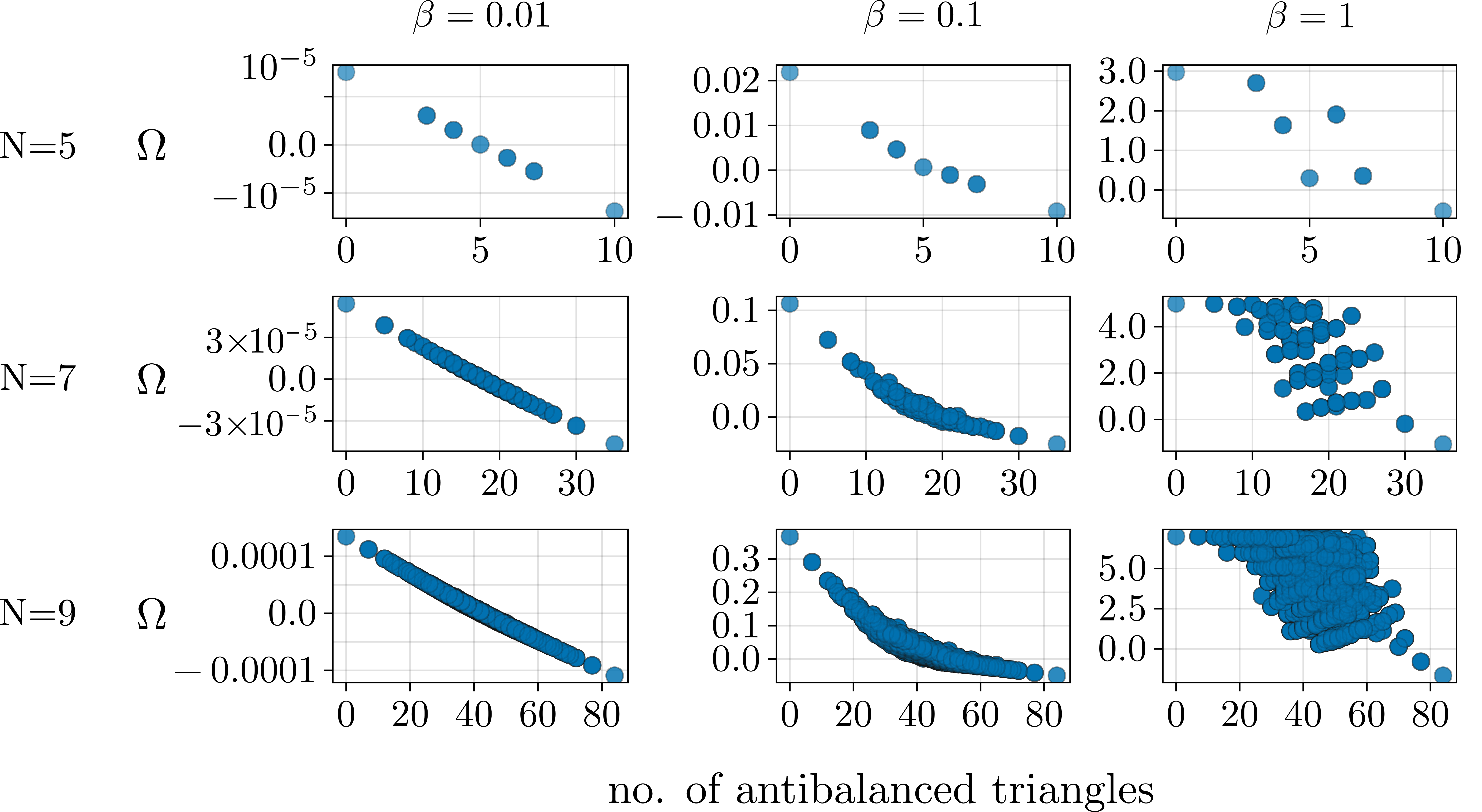}
    \caption{We show the O-information versus the number of antibalanced triangles in the Ising model for $N=5,7,9$ and $\beta=0.01, 0.1, 1$, for all possible signed coupling configurations $J_{ij}$. As for the case of $N=6$, $\beta=0.25$ reported in the main text (Fig. $3$), antibalanced coupling motifs are necessary for synergy-dominance, and the most synergistic structures have the the highest number of antibalanced triangles. Additionally, here we see that at least $\binom{N}{3}/2 + 1$ antibalanced triangles are needed for synergy-dominance, i.e., more than half of the triangles need to be antibalanced for the system to be synergy-dominated.}
    \label{fig:Ising-extra}
\end{figure}

\textit{Model parameters.} As for the OU process study (see Sec.~\ref{sec:SM-OU-process-methodology}), we computed the O-information of the Ising model described above for each possible coupling structure ($N<10$). For simplicity, we fixed the couplings at $|J_{jk}|=1$, such that the only free parameter is the inverse temperature $\beta$. In Fig. \ref{fig:Ising-extra}, we show additional results for $N\in\{5, 7, 9\}$ and for $\beta\in\{0.01, 0.1, 1\}$. All computations here are exact (i.e., no sampling).

\subsection{System of Stuart-Landau oscillators}
\textit{Oscillator model.} The intrinsic (uncoupled) dynamics of a Stuart-Landau (SL) oscillator is described by the normal form of a supercritical Hopf bifurcation
\begin{equation}
    \frac{\mathrm{d}Z_{j}}{\mathrm{d}t}=Z_j (a + i\omega - |Z_j|^2) + \beta\eta_1 + i\beta\eta_2 \label{eqn:SL_intrinsic-dynamics}
\end{equation}
where $Z_j=Z_1 + iZ_2$ denotes the state of oscillator $j$, $a$ is the bifurcation parameter, and $\omega=2\pi f$ is the angular frequency ($f_j$ denotes the intrinsic oscillator's frequency). At $a=0$, the system undergoes a supercritical bifurcation. In the subcritical regime ($a<0$), if $a$ is sufficiently close to $0$, the system exhibits damped oscillations induced by the complex Gaussian noise $\eta_1 + i\eta_2$ with standard deviation $\beta$. When $a\ll0$, the system is overdamped and oscillations decay very quickly. In the supercritical regime ($a>0$) the system exhibits a stable limit cycle with frequency $f_j$.

\textit{Coupled oscillators model.} The coupled system of SL oscillators of size $N$ can be written as:
\begin{equation}
    \frac{\mathrm{d}Z_{j}}{\mathrm{d}t}=Z_j (a + i\omega - |Z_i|^2) + K\sum_{k\neq{j}}A_{jk}\big(Z_k - Z_{j}\big) + \beta\eta_1 + i\beta\eta_2,\label{eqn:SL-model}
\end{equation}
where $K$ is the global coupling parameter and $A$ is the interaction matrix. Again, we consider all-to-all interactions and allow both positive and negative interactions $A_{jk}=\pm1$. In the results we present here and the main text, all oscillators are identical: same intrinsic frequency and same bifurcation parameter.

\begin{figure}[h]
    \centering
    \includegraphics[width=.8\linewidth]{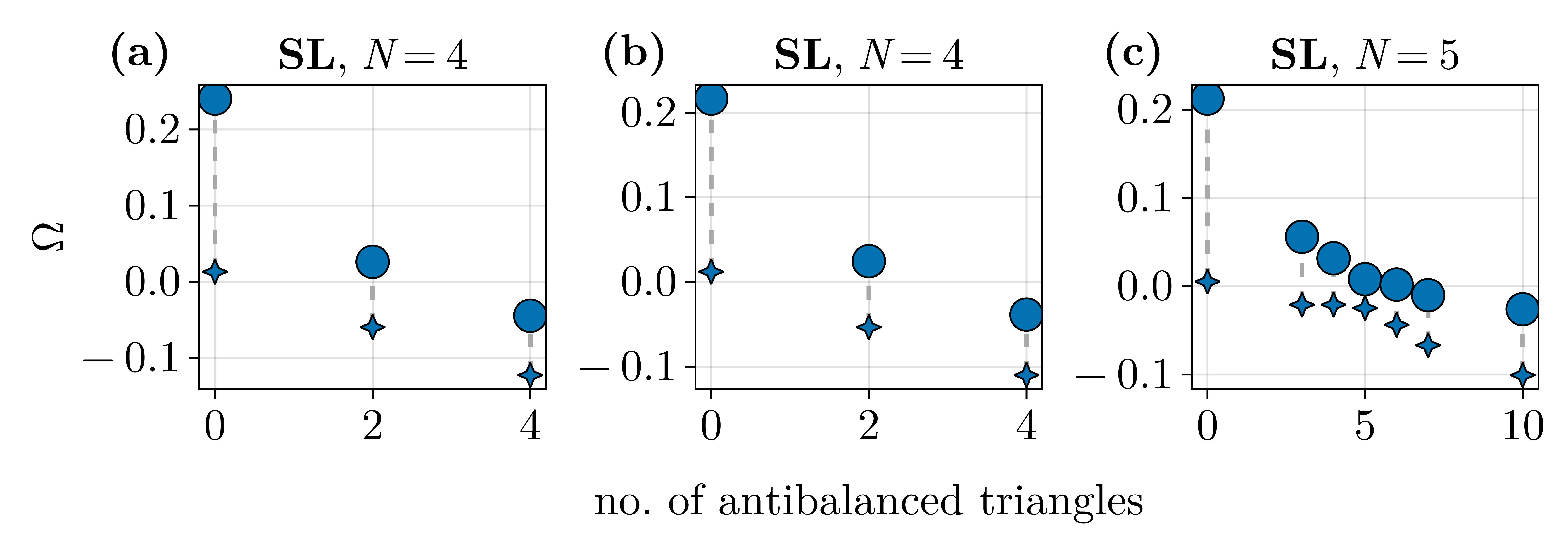}
    \caption{We show the O-information versus the number of antibalanced triangles in networks of SL oscillators for all possible signed coupling configurations $A$. We use global coupling $K_0=4$ and (\textbf{a}) $N=4$ and filtered data, (\textbf{b}) $N=4$ and unfiltered data, (\textbf{c}) $N=5$ filtered data. As for the case of $N=6$, reported in the main text (Fig. $3$), antibalanced coupling motifs are necessary for synergy-dominance, and the most synergistic structures have the the highest number of antibalanced triangles.}
    \label{fig:SL-extra}
\end{figure}

\textit{Information theoretic quantities for the SL model.} Using the simulated real parts $\operatorname{Re}Z_j(t)$, we compute the O-information using the \texttt{HOI} Python package~\cite{Neri2024}. For the entropy estimators we used the Gaussian copula method implemented in \texttt{HOI}, as it is commonly used to study fMRI and EEG data~\cite{Ince2016}. Similarly to~\cite{Cabral2022, Kringelbach2023, Luppi2024Competitive}, signals were band-pass filtered with a $4$th-order Butterworth filter ($35$–$45\mathrm{Hz}$) before computing the information theoretic quantities. In Fig.~\ref{fig:SL-extra} we show additional results for $N=4,5$, demonstrating that our conclusions hold when different parameters are used, or if unfiltered data is used to compute the information-theoretic quantities.

\textit{Model implementation and parameters.} We adapted the code from \href{https://github.com/fcast7/Hopf_Delay_Toolbox}{https://github.com/fcast7/Hopf\_Delay\_Toolbox} into Julia~\cite{github_repo_synergistic_motifs}. As in \cite{Cabral2022}, we used: Euler step of size $10^{-4}$; oscillator intrinsic frequency $f=40$Hz; Gaussian noise standard deviation $\beta=0.001$; bifurcation parameter $a=-5.0$; and random initial conditions. The results presented here and in the main text were obtained using a global coupling (i.e., the same for each node) that depends on the configuration of the system: $K=K_0/(N+2b(A))$, where $b(A)$ is the number of negative edges in the system, and $K_0$ was set to be equal to $4$ for all experiments. The parameter $b(A)$ is crucial for comparing the dynamics of systems with a different number of negative edges as, with the addition of negative edges, quenching behaviours~\cite{Koseska2013} such as amplitude death~\cite{Hens2013} may appear. All simulations are run for $5\times10^5$ steps (equiv. $50$ seconds), and the first $10^5$ steps are discarded.

\subsection{Empirical datasets}
\textit{fMRI dataset.}---We analysed the openly available neuroimaging dataset by Rieck \textit{et al.}~\cite{Rieck2021}. To produce the figure in the main text, we used the file \texttt{Power229\_10\_sub-001\_rest.txt} which contains a single participant’s resting‑state functional connectivity matrix. Using the code we provide in~\cite{github_repo_synergistic_motifs}, qualitatively equivalent results can be obtained using other participants' resting state data, other parcellation schemes, and using the average correlation matrix across participants. We report in Table S1 the relevant statistics for the spectral radius of the $N$-plets studied in the main text.

Since this dataset was used in~\cite{Saberi2024} to investigate the functional organisation of brain networks across tasks using SBT (i.e., focusing on $U$) here we show that an analogous analysis can be performed using the O-information. Specifically, we compute the average O-information across triplets and show our results are consistent with those from $U$.

\begin{figure}
    \centering
    \includegraphics[width=1\linewidth]{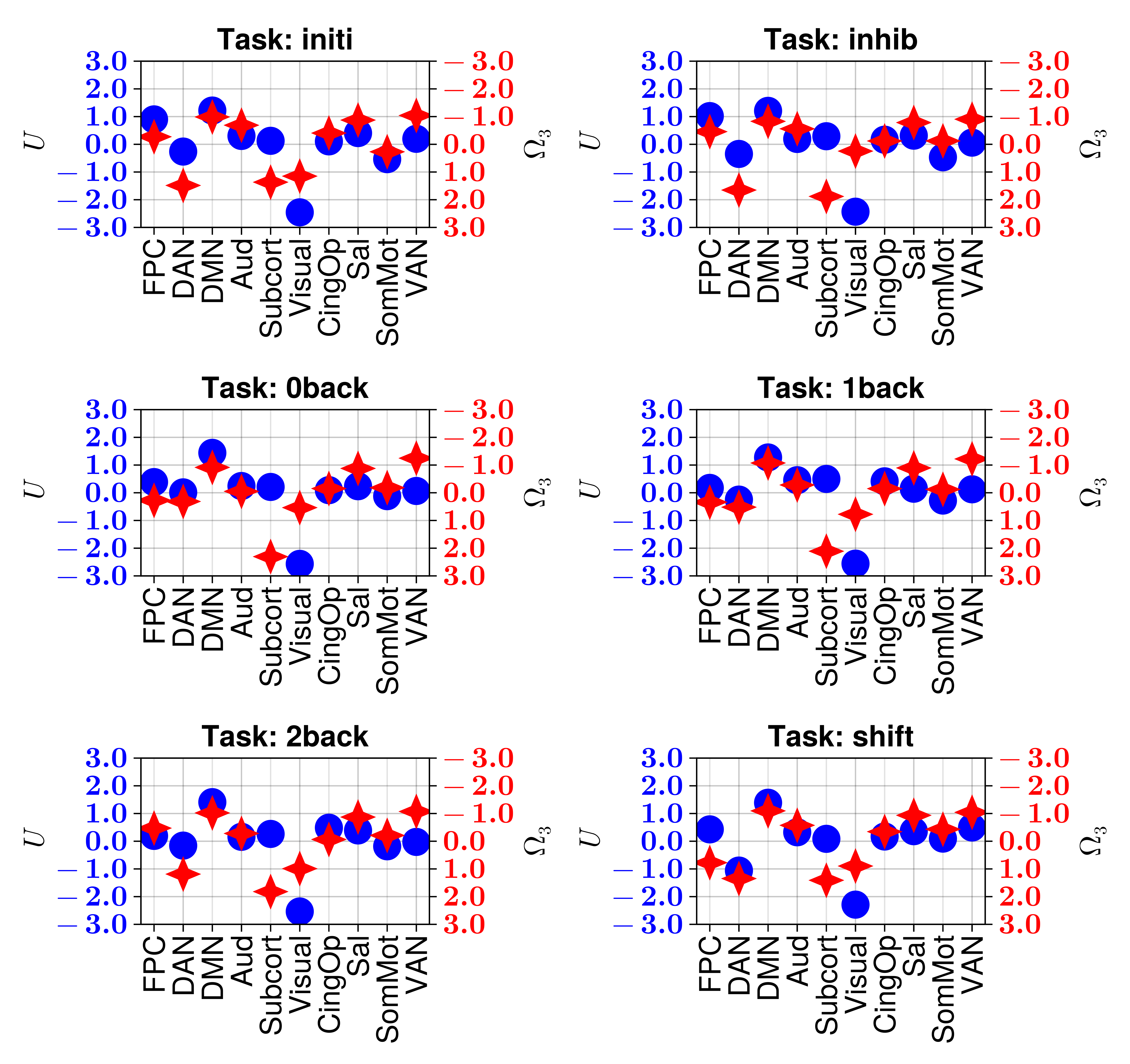}
    \caption{Balance-energy $U$ and $\Omega$ of each canonical networks across tasks. Markers depict median values across individuals.}
    \label{fig:U_vs_O_tasks}
\end{figure}

In practice, we load the task-specific correlation matrices provided by Rieck \textit{et al.}~\cite{Rieck2021}, group the nodes into canonical functional networks according to the given parcellation (see~\cite{Rieck2021}), and compute both the structural balance-energy $U$ and the average O-information across all node triplets, $\Omega_3$. We repeat this for each task, functional network grouping, and each individual. Across all task and network combinations ($7$ tasks and $10$ canonical networks), our results (Fig.~\ref{fig:U_vs_O_tasks}) show that the structural balance energy $U$ and the average triplet O-information $\Omega_3$ are strongly, negatively correlated (Pearson $r = -0.62$, $p < 10^{-10}$). This analysis exemplifies how $U$ and $\Omega$ capture very closely related aspects of the functional organisation of brain dynamics, while providing (often complementary) distinct interpretations (e.g., in~\cite{Saberi2024} high $U$ reflects frustration, tension, and high energy cost, whereas in~\cite{Luppi2024B} synergy reflects cooperation, high-order functional organisation, and efficiency).

\textit{Foreign-exchange (FX) logarithm returns}---We analysed the multivariate timeseries collected in~\cite{CliffDataset} (see also the study~\cite{Cliff2023}). For the figure in the main text, we used the file \texttt{ForEx\_N-886\_P-9\_2010-01-01-lr}. Using the code we provide in~\cite{github_repo_synergistic_motifs}, the same analysis can be repeated for all multivariate time series datasets collected in~\cite{CliffDataset}. As for the fMRI data, we report in Table S2 the relevant statistics for the spectral radius of the FX $N$-plets studied in the main text.

\textit{Information theoretic quantities}---For both dataset (fMRI's correlation matrix and the FX logarithmic returns timeseries) we used the openly available Python package \texttt{THOI} \cite{THOI} to identify the most synergistic and most redundant $N$-plets. For the financial data, all $N$-plets of size $N=3,4,5,6,7$ were computed exhaustively. For the fMRI data instead, we used the greedy algorithm to identify the $10000$ most synergistic and the $10000$ most redundant $N$-plets of size $N=3,4,5,6,7$.

\begin{figure}
    \centering
    \includegraphics[width=1\linewidth]{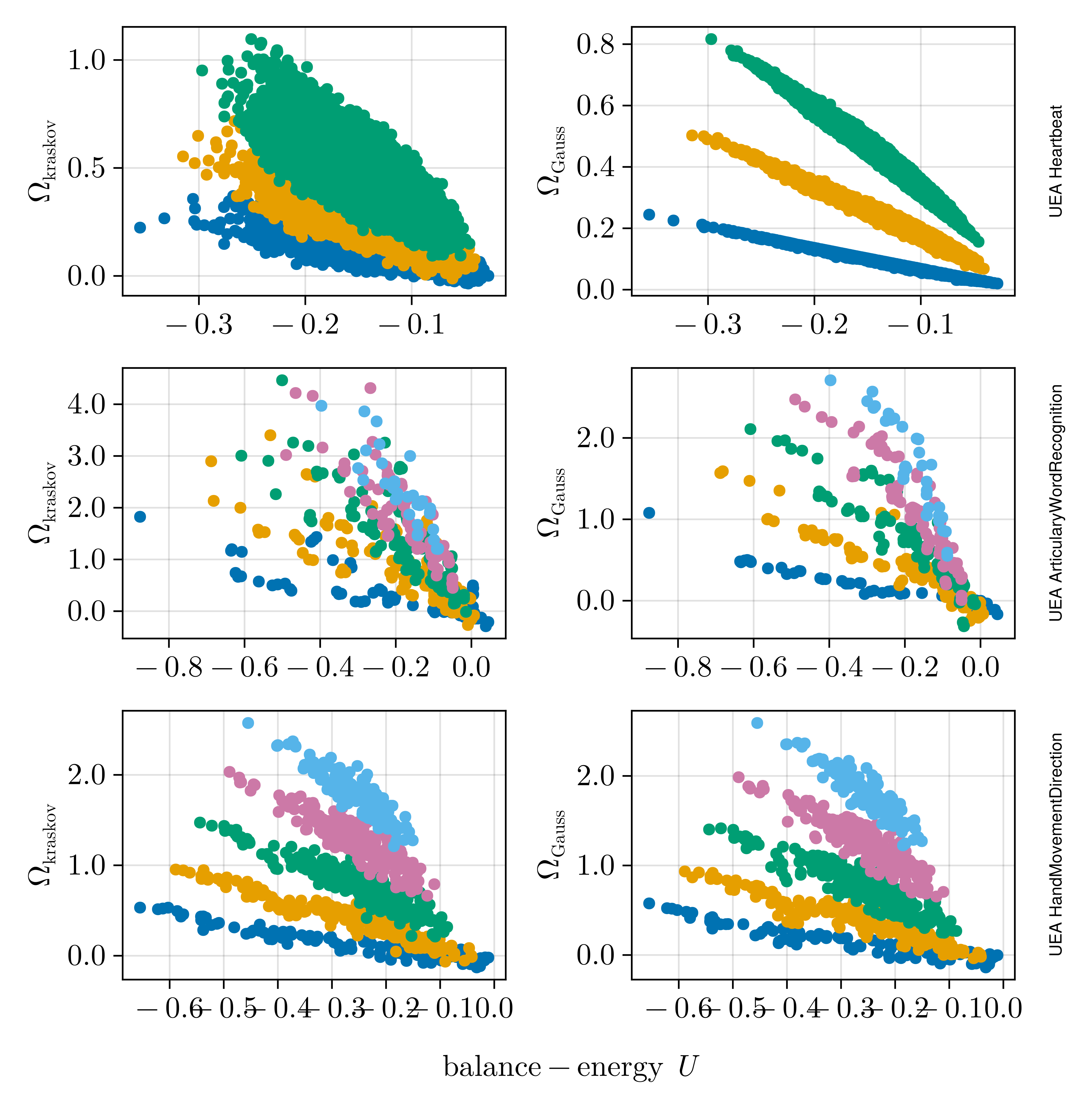}
    \caption{\textbf{Datasets from medicine: Kraskov O-information (left) Gaussian O-information (right)}. For more information on the datasets, see Supplementary Information (SI) accompanying~\cite{Cliff2023}}
    \label{fig:kraskov_vs_gaussian1}
\end{figure}

\begin{figure}
    \centering
    \includegraphics[width=1\linewidth]{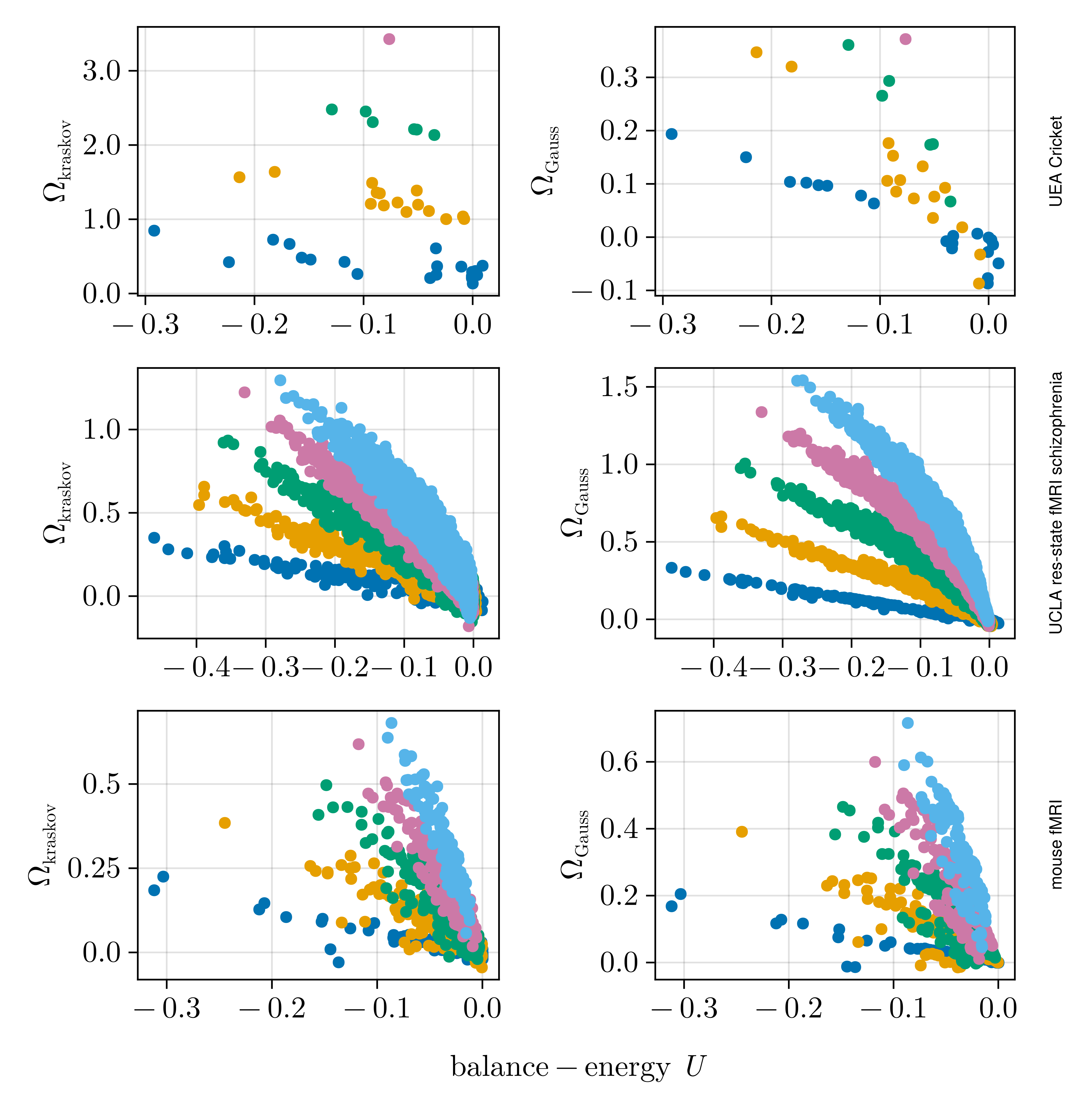}
    \caption{\textbf{Datasets from physiology and mice fMRI data: Kraskov O-information (left) Gaussian O-information (right)}. For more information on the dataset, see SI accompanying~\cite{Cliff2023}}
    \label{fig:kraskov_vs_gaussian2}
\end{figure}

\begin{figure}
    \centering
    \includegraphics[width=1\linewidth]{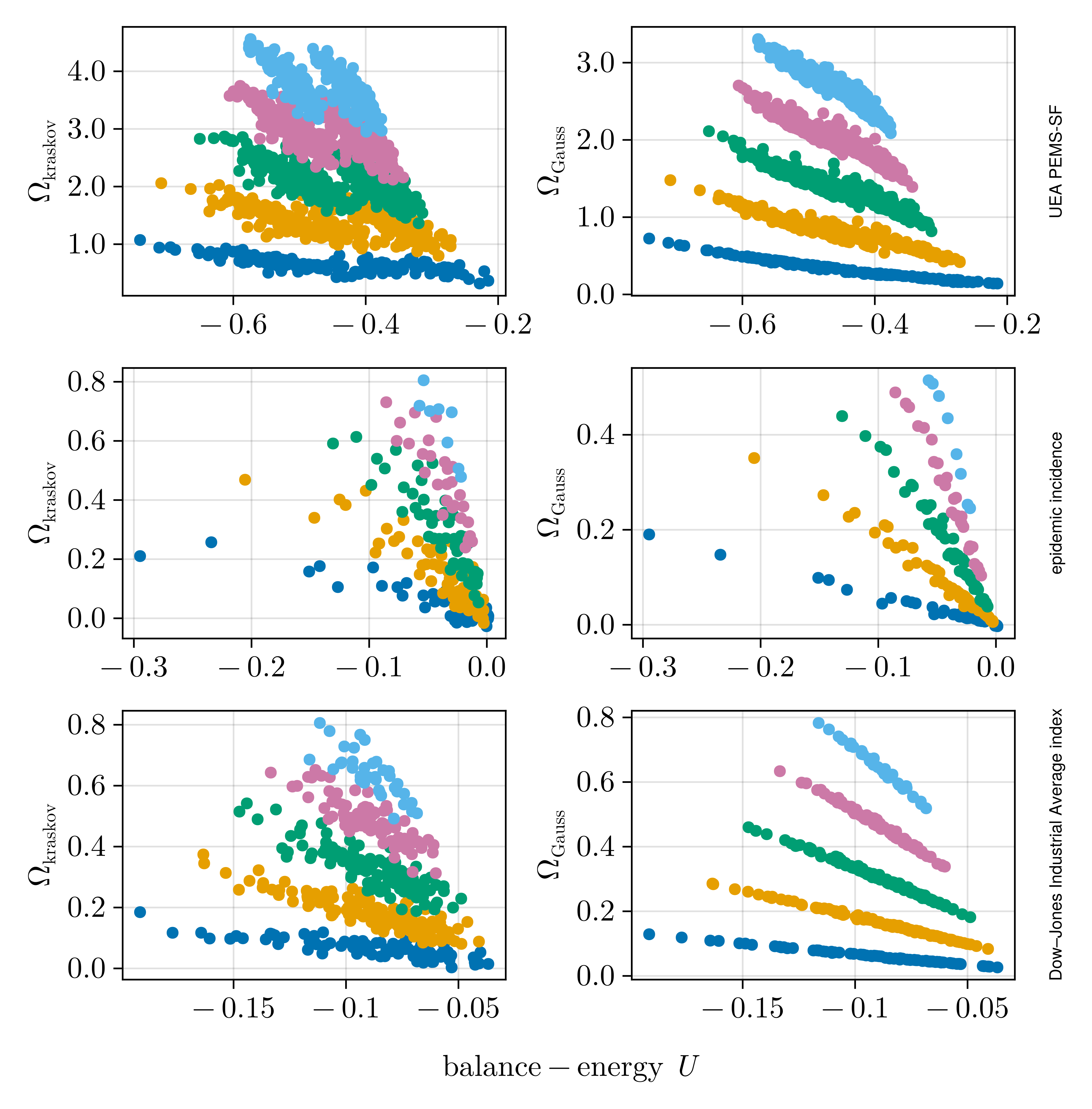}
    \caption{\textbf{Car lanes occupancy rates, epidemic, and financial data: Kraskov O-information (left) Gaussian O-information (right)}. For more information on the dataset, see SI accompanying~\cite{Cliff2023}}
    \label{fig:kraskov_vs_gaussian3}
\end{figure}

\textit{Kraskov vs Gaussian and further analyses}---To test the robustness of our conclusions and the impact of Gaussian assumptions, we performed an additional study comparing estimates of the O-information using both Gaussian copula estimators (implemented with the \texttt{THOI} Python package) and nonparametric Kraskov $k$-nearest-neighbour estimators (implemented using the \texttt{JIDT} toolkit~\cite{Lizier2014}). Notably, both estimates of the O-information can be sensitive to high-order interactions. In Figs.~\ref{fig:kraskov_vs_gaussian1}–\ref{fig:kraskov_vs_gaussian3}, we report examples from medicine, physiology, finance, and epidemiology. Across all datasets, Kraskov-based and Gaussian-copula-based O-information estimates were highly consistent, both showing a systematic relationship with the structural balance-energy $U$ (the latter computed using the correlation matrix of the time series estimated using the Gaussian copula method). Additional information on the datasets and preprocessing can be found in the Supplementary Information (SI) accompanying~\cite{Cliff2023}.

\begin{table}[t]
\centering
\label{tab:critical_vs_subcritical_emp_data_fMRI}
\begin{tabular}{c c c c c c c}
\hline
$N$ & type & $n_{\text{studied}}$ & $p_{\text{strong}}$ (\%) & $\rho_{\min}$ & $\rho_{\text{mean}}$ & $\rho_{\max}$ \\
\hline
3  & synergistic & 10000 & 0.37  & 0.525 & 0.654 & 1.178 \\
3  & redundant   & 10000 & 100.0 & 1.159 & 1.294 & 1.783 \\
5  & synergistic & 10000 & 3.93  & 0.591 & 0.779 & 1.745 \\
5  & redundant   & 10000 & 100.0 & 2.079 & 2.873 & 3.401 \\
7  & synergistic & 10000 & 18.25 & 0.637 & 0.881 & 1.820 \\
7  & redundant   & 10000 & 100.0 & 2.737 & 4.395 & 4.875 \\
10 & synergistic & 10000 & 64.0  & 0.658 & 1.085 & 2.021 \\
10 & redundant   & 10000 & 100.0 & 3.966 & 6.551 & 6.920  \\
\hline
\end{tabular}
\caption{\textbf{Summary statistics of the spectral radius $\rho(W)$ for synergistic and redundant $N$-plets from the resting-state fMRI dataset studied in the main text (Fig.~$4$)}.
An $N$-plet is in the strong-interactions regime if $\rho(W)\geq1$. For each order $N$, we studied the $10000$ most synergistic and most redundant $N$-plets and report the percentage of $N$-plets with $\rho(W)\geq1$, i.e., $p_{\text{strong}}$ as well as the minimum, mean, and maximum spectral radius.}
\end{table}

\begin{table}[t]
\centering
\label{tab:critical_vs_subcritical_emp_data_FX}
\begin{tabular}{c c c c c c c}
\hline
$N$ & $n_{\text{studied}}$ & $p_{\text{strong}}$ (\%) & $\rho_{\min}$ & $\rho_{\text{mean}}$ & $\rho_{\max}$ \\
\hline
3 & 84  & 36.905 & 0.125 & 0.796 & 1.616 \\
4 & 126 & 76.984 & 0.415 & 1.236 & 2.108 \\
5 & 126 & 97.619 & 0.883 & 1.683 & 2.571 \\
6 & 84  & 100.0  & 1.363 & 2.132 & 2.994 \\
7 & 36  & 100.0  & 2.045 & 2.582 & 3.439 \\
\hline
\end{tabular}
\caption{\textbf{Summary statistics of the spectral radius $\rho(W)$ for synergistic and redundant $N$-plets from the FX dataset studied in the main text (Fig.~$4$)}.
For each order $N$, we studied all possible $N$-plets ($n_{\text{studied}}=\binom{9}{N}$ since there are $9$ timeseries here) and report the percentage of $N$-plets with $\rho(W)\geq1$, i.e., $p_{\text{strong}}$ as well as the minimum, mean, and maximum spectral radius.}
\end{table}

\putbib
\end{bibunit}


\end{document}